\documentclass[english]{IEEEtran}
\usepackage[T1]{fontenc}
\usepackage[latin9]{inputenc}
\usepackage{amsmath}
\usepackage{amsfonts}
\usepackage{amssymb}
\usepackage{graphicx}
\usepackage{subfigure}
\usepackage{esint}

\makeatletter
\newtheorem{prop}{Proposition}
\newtheorem{example}{Example}

\makeatother

\begin{document}

\title{Time Delay Estimation from Low Rate Samples: A Union of Subspaces Approach}
\author{Kfir Gedalyahu and Yonina C.~Eldar,~\IEEEmembership{Senior~Member,~IEEE}
\thanks{Department of Electrical Engineering,
Technion---Israel Institute of Technology, Haifa 32000, Israel.
Phone: +972-4-8293256, fax: +972-4-8295757, E-mail:
\{kfirge@techunix,yonina@ee\}.technion.ac.il. This work was
supported in part by the Israel Science Foundation under Grant no.
1081/07 and by the European Commission in the framework of the FP7
Network of Excellence in Wireless COMmunications NEWCOM++
(contract no. 216715).}}

\maketitle
\begin{abstract}
Time delay estimation arises in many applications in which a
multipath medium has to be identified from pulses transmitted
through the channel. Various approaches have been proposed in the
literature to identify time delays introduced by multipath
environments. However, these methods either operate on the analog
received signal, or require sampling at the Nyquist rate of
the transmitted pulse. In this paper, our goal is to
develop a unified approach to time delay estimation from low rate
samples of the output of a multipath channel. Our methods result
in perfect recovery of the multipath delays from samples of the
channel output at the lowest possible rate, even in the presence
of overlapping transmitted pulses. This rate depends only on the
number of multipath components and the transmission rate, but not
on the bandwidth of the probing signal. In addition, our
development allows for a variety of different sampling methods. By
properly manipulating the low-rate samples, we show that the time
delays can be recovered using the well-known ESPRIT algorithm.
Combining results from sampling theory with those obtained in the
context of direction of arrival estimation methods, we develop sufficient
conditions on the transmitted pulse and the sampling functions in order to
ensure perfect recovery of the channel parameters at the minimal
possible rate.
Our results can be viewed in a broader context, as a sampling theorem for analog
signals defined over an infinite union of subspaces.
\end{abstract}

\section{\label{sec:Intro}Introduction}

Time delay estimation is an important signal processing problem,
arising in various applications such as radar \cite{quazi1981otd},
underwater acoustics \cite{urick1983pus}, wireless communications
\cite{turin1980iss}, and more. In a typical scenario, pulses with
a priori known shape are transmitted through a multipath medium,
which consists of several propagation paths. As a result the
received signal is composed of delayed and weighted replicas of
the transmitted pulses. In order to identify the medium, the time
delay and gain coefficient of each multipath component has to be
estimated from the received signal.

In this paper we consider recovery of the parameters defining such
a multipath medium from samples of the channel output.
Specifically, we assume that pulses with known shape are
transmitted at a constant rate through the medium, and our aim is
to recover the time delays and time-varying gain coefficients of
each multipath component, from samples of the received signal. We
focus on the sampling stage, and derive methods that ensure
perfect recovery from samples of the channel output at the minimal
possible rate. The proposed sampling schemes are flexible, so that
a variety of different sampling techniques can be accommodated.
Our main contribution is the development of efficient sampling
schemes for the received signal. The resulting sampling rate is
generally much lower than the traditional Nyquist rate of the
transmitted pulses, and depends only on the number of multipath
components and the transmission rate, but not on the bandwidth of
the transmitted pulse.
This can lead to significant sampling rate reduction, comparing to the Nyquist rate, in applications where only small number of propagation paths exists, or when the bandwidth of the transmitted pulse is relatively high.
This reduction is desirable for practical implementation.
Sampling at lower rates allows for analog-to-digital converters (ADCs) that are more precise (i.e. use more bits), and with lower power consumption. In addition, lowering the sampling rate can reduce the load on both hardware and further digital processing units.

A classical solution for the time delay estimation problem is
based on correlation between the received signal and the
transmitted pulse \cite{quazi1981otd}. However, the time
resolution of this method is limited by the inverse of the
transmitted pulse bandwidth. Therefore, this technique is
effective only when the multipath components are well separated in
time, or when only one component is present. This approach was
originally motivated in the analog domain, where the entire analog
correlation is computed. Performing the correlation in the digital
domain requires samples of the data at a high sampling rate, in
order to approximate the continuous correlation.

 In order to
resolve closely spaced multipath components, various
superresolution estimation algorithms were proposed. In
\cite{bruckstein1985roe,pallas1991ahr}, the MUSIC \cite{1143830}
method was applied in the time domain in order to estimate the
time delay of each multipath component. Hou and Wu
\cite{ziqiang1982nmh} were the first to convert the time
estimation problem to model-based sinusoidal parameter estimation,
and used an autoregressive method in order to estimate the model's
parameters. Other works, such as
\cite{pallas1991ahr,saarnisaari1997tet,4303294}, relied on the
same principle, but different estimation algorithms were used:
Tufts-Kumaresan SVD algorithm \cite{kumaresan1983eaa}, TLS-ESPRIT
method \cite{32276} and a modification of MUSIC, respectively.

In the above superresolution approaches, the sampling stage was
typically not directly addressed. Most of these works rely on
uniform pointwise samples of the received signal, at a high
sampling rate. In \cite{ziqiang1982nmh} and \cite{4303294} the required sampling rate
is the Nyquist rate of the transmitted pulse. Since often the
pulses are chosen to have small time support, the bandwidth can be
quite large, corresponding to a high Nyquist rate. In
\cite{saarnisaari1997tet} and in the frequency domain algorithm
proposed in \cite{pallas1991ahr}, the sampling of the received
analog signal is not mentioned explicitly. Since these algorithms
involve operations in the analog frequency domain, effectively
they also require sampling at the Nyquist rate. The time domain
algorithms proposed in \cite{bruckstein1985roe} and
\cite{pallas1991ahr} can theoretically recover the time delays
from a low sampling rate, which depends on the number of multipath
components. However, the sampling considerations were not
directly addressed in these works, and no concrete conditions on
the transmitted pulse and the samples were given, in order to
ensure unique recovery of the delays from the samples.

Besides the sampling stage which is not studied in previous works,
another assumption underlying all the methods above is that the
receiver has access to a single experiment
(\cite{ziqiang1982nmh,saarnisaari1997tet,4303294}) or multiple
non-overlapping experiments
(\cite{bruckstein1985roe,pallas1991ahr,4303294}) on the channel.
In each experiment a pulse is transmitted through the multipath
medium, and it is required that all the returns vanish before the
next experiment is conducted. This imposes the constraint that the
transmitted pulse is sufficiently time limited, which can be
problematic in certain scenarios. For example, in wireless
communications, modulated pulses are transmitted at a constant
symbol rate through the medium. In this case, we cannot consider
the observed signal over one symbol period as an independent
experiment, since it will generally be affected by reflections
caused by adjacent symbols.

In Section~\ref{sec:Signal-Model-and} we propose a general signal
model, that can describe the received signal from a time-varying
multipath medium. An advantage of our model is that it does not
require the assumption of non-overlapping experiments, and allows
for general pulse shapes. We then formulate the medium
identification problem as a sampling problem, in which the set of
parameters defining the medium have to be recovered from samples
of the received signal at the lowest possible rate. To this end we
develop a general sampling scheme, which consists of filtering the
received signal with a bank of $p$ sampling filters and uniformly
sampling their outputs. This class of sampling methods is common
in sampling theory \cite{EM08} and can accommodate a wide variety
of sampling techniques, including pointwise uniform sampling.
Given $K$ multipath components, we show that at least $2K$
sampling filters are required in order to perfectly recover the
time delays. We then develop explicit sampling strategies that
achieve this minimal rate. In particular, we derive sufficient
conditions on the transmitted pulse and the choice of sampling
filters, which guarantee unique recovery of the channel
parameters.

In order to recover the channel parameters from the given samples
we combine results from standard sampling theory, with those of
direction of arrival (DOA) algorithms
\cite{1143830,32276,Johnson1993,krim1996tda}. Specifically, by
appropriate manipulation of the sampling sequences, we show that
we can formulate our problem within the framework of DOA methods.
We then rely on the estimation of signal parameters via rotational
invariance techniques (ESPRIT) \cite{32276}, developed in that
context. The unknown delays are recovered from the samples by
first applying a digital correction filter bank, and then applying
the ESPRIT algorithm. Once the time delays are identified, the
gain coefficients are recovered using standard sampling tools.

The sampling schemes we develop for the channel identification
problem treated in this paper, can be viewed in the broader
context of sampling theory. In Section~\ref{sec:Related-Works} we
discuss in detail the relationship between our problem and
previous related setups treated in the sampling literature:
sampling signals from a union of subspaces
\cite{4483755,BD09},
compressed sensing of analog signals
\cite{location,ME07,Mishali,EM082,4553693,E082}, and finite-rate of
innovation (FRI) sampling \cite{1003065,4156380}. The results we
develop here can be viewed as a special case of analog compressed
sensing over an infinite union of spaces, and therefore extends
previous work, which focused on finite unions, to a broader
setting. In comparison with FRI techniques, our approach allows
for lower sampling rates and at a lower computational cost.
Furthermore, we do not need to impose stringent conditions on the
pulse shapes, as required by FRI methods.

This paper is organized as follows. In Section
\ref{sec:Signal-Model-and}, we describe our signal model. A
general sampling scheme of the received signal is proposed in
Section \ref{sec:Sampling-Scheme}. Section
\ref{sec:Unknown-Delays-Recovery} describes the recovery of the
unknown delays from the samples, and provides sufficient
conditions ensuring a unique recovery. Relation to previous work
is discussed in detail in Section \ref{sec:Related-Works}.
Section~\ref{sec:Application} describes an application example of
channel identification in wireless communications. Numerical
experiments are described in Section
\ref{sec:Numerical-Experiments}.

\section{\label{sec:Signal-Model-and}Signal Model and Problem formulation}

\subsection{Notations and Definitions}

Matrices and vectors are denoted by bold font, with lowercase
letters corresponding to vectors and uppercase letters to
matrices. The $n$th element of a vector $\mathbf{a}$ is written as
$\mathbf{a}_{n}$, and the $ij$th element of a matrix $\mathbf{A}$
is denoted by $\mathbf{A}_{ij}$. Superscripts
$\left(\cdot\right)^{*}$, $\left(\cdot\right)^{T}$ and
$\left(\cdot\right)^{H}$ represent complex conjugation,
transposition and conjugate transposition, respectively. The
Moore-Penrose pseudo-inverse of a matrix $\mathbf{A}$ is written
as  $\mathbf{A}^{\dagger}$. The identity matrix of size $n$ is
denoted by $\mathbf{I}_{n}$.

The Fourier transform of a continuous-time signal
$x\left(t\right)\in L_{2}$ is defined by
$X\left(\omega\right)=\int_{-\infty}^{\infty}x\left(t\right)e^{-j\omega t}d\omega$,
and
\begin{equation}
\left\langle x\left(t\right),y\left(t\right)\right\rangle
=\int_{-\infty}^{\infty}x\left(t\right)y^{*}\left(t\right)dt,
\end{equation}
denotes the inner product between two continuous-time signals. The
discrete-time Fourier transform (DTFT) of a sequence
$a\left[n\right]\in\ell_{2}$ is defined by
\begin{equation}
A\left(e^{j\omega
T}\right)=\sum_{n\in\mathbb{Z}}a\left[n\right]e^{-j\omega nT},
\end{equation}
and is periodic with period $2\pi/T$.

\subsection{Signal model}

 We consider the class of signals that can be written in the form
\begin{equation}
x\left(t\right)=\sum_{k=1}^{K}\sum_{n\in\mathbb{Z}}a_{k}\left[n\right]g\left(t-t_{k}-nT\right),
\label{eq:x(t)}\end{equation} where $\tau=\left\{ t_{k}\right\}
_{k=1}^{K}$ is a set of $K$ distinct unknown time delays in the
continuous interval $\left[0,T\right)$, $a_{k}\left[n\right]$ is
an arbitrary sequence in $\ell_{2}$, and $g\left(t\right)\in
L_{2}$ is a known function. Each signal from this class can
describe the propagation of a pulse with known shape
$g\left(t\right)$ which is transmitted at a constant rate of $1/T$
through a medium consisting of $K$ paths. Each path has a constant
delay $t_{k}$, and a time-varying gain, which is described by the
sequence $a_{k}\left[n\right]$. In cases where the transmitted
pulses are amplitude modulated, the sequence $a_{k}\left[n\right]$
describes the multiplication between the pulse amplitude and the
gain coefficient of each path. In Section \ref{sec:Application} we
will discuss more thoroughly an example of a communication signal
transmitted through a multipath time-varying channel.

Our problem is to determine the delays $\tau$ and the gains
$a_k[n]$ from samples of the received signal $x(t)$, at the
minimal possible rate. Since these parameters uniquely define
$x(t)$, our channel identification problem is equivalent to
developing efficient sampling schemes for signals of the form
(\ref{eq:x(t)}), allowing perfect reconstruction of the signal
from its samples.

The model (\ref{eq:x(t)}) is more general than that described in
previous work
\cite{ziqiang1982nmh,bruckstein1985roe,pallas1991ahr,saarnisaari1997tet,4303294},
which is based on single or multiple experiments on the medium. In
each experiment the received signal is observed over a finite time
window, which is synchronized to the transmission time of the
pulse. More precisely, the received signal in the $n$th experiment
is given by\begin{equation} \label{eq:prev}
x_{n}\left(t\right)=\sum_{k=1}^{K}a_{kn}g\left(t-t_{k}\right),\quad
t\in\left[0,T\right),\end{equation} where $t_{k}$ is the delay of
the $k$th multipath component which is constant in all the
experiments, and $a_{kn}$ is the gain coefficient of the $k$th
multipath component at the $n$th experiment, which is generally
varying from one experiment to another. This model can be seen as
a special case of \eqref{eq:x(t)} with additional constraints on
the pulse $g\left(t\right)$ and the transmission rate $1/T$.
Indeed, we  can write the received signal on the $n$th experiment
as
\begin{equation} x_{n}\left(t\right)=x\left(t-nT\right)\quad
t\in\left[0,T\right),\end{equation} if we require that
\begin{equation} g\left(t-t_{k}-nT\right)=0\textrm{ for all
}n\neq0.\end{equation} This requirement means that the pulse
$g\left(t\right)$ has finite time support, and that the repetition
period of the pulses $T$, is long enough such that all the
reflections from one pulse vanish before the next pulse is
transmitted. On the other hand, our signal model does not require
these constraints, so that it can support infinite length pulses
and allows interference between experiments.

\section{\label{sec:Sampling-Scheme}Sampling Scheme}

\subsection{\label{subsec:Known}Known delays}

Before we treat our sampling problem of signals of the form
(\ref{eq:x(t)}), we first consider a simpler setting in which the
delays $t_k$ are known. In this case our signal model is a special
case of the more general class of signals that lie in a
shift-invariant (SI) subspace. For such signals classes, there are
well known sampling schemes that guarantee perfect recovery at the
minimal possible rate \cite{EM08,location,AG01}. Below, we review
the main results in this setting which will serve as a basis for
our development in the case in which the delays are unknown.

A finitely-generated SI subspace of $L_{2}$ is defined as
\cite{deboor1994sfg,geronimo1994ffa,christensen2005gsi}\begin{equation}
\mathcal{A}=\left\{
x\left(t\right)=\sum_{k=1}^{K}\sum_{n\in\mathbb{Z}}a_{k}\left[n\right]g_{k}\left(t-nT\right):a_{k}\left[n\right]\in\ell_{2}\right\}
.\label{eq:finitelygenerated}\end{equation} The functions
$g_{k}\left(t\right)$ are referred to as the generators of
$\mathcal{A}$. In order to guarantee a unique stable
representation of any signal $x\left(t\right)\in\mathcal{A}$ by
coefficients $a_{k}\left[n\right]$, the generators
$g_{k}\left(t\right)$ are typically chosen to form a Riesz basis
of $\mathcal{A}$ \cite{deboor1994sfg,geronimo1994ffa}. Clearly,
the signal model in \eqref{eq:x(t)} is a special case of
\eqref{eq:finitelygenerated} with generators $g_k(t)$, obtained
from $K$ delayed versions of $g\left(t\right)$:\begin{equation}
g_{k}\left(t\right)=g\left(t-t_{k}\right),\quad1\leq k\leq
K.\end{equation}

\begin{figure}
\begin{center}
\includegraphics[scale=0.5]{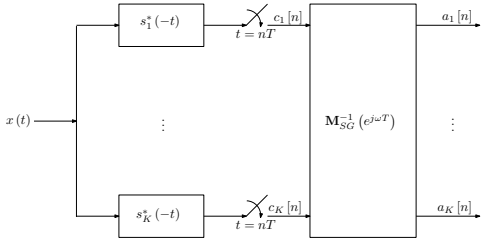}

\caption{\label{fig:SI}Sampling and reconstruction scheme for the case of
known delays}

\end{center}
\end{figure}

One way to sample a signal of the form
\eqref{eq:finitelygenerated} is to use $K$ parallel sampling
channels \cite{location}. In each channel the signal is first
filtered with an impulse response $s_{\ell}^{*}\left(-t\right)$,
and then sampled uniformly at times $t=nT$ to produce the sampling
sequence $c_{\ell}\left[n\right]$, as depicted in the left-hand
side of Fig.~\ref{fig:SI}. The sampling sequence at the output of
the $\ell$th channel can be written as \begin{equation}
\label{eq:cn} c_{\ell}\left[n\right]=\left\langle
x\left(t\right),s_{\ell}\left(t-nT\right)\right\rangle
,\quad1\leq\ell\leq K.\end{equation} By analyzing the DTFTs
$C_{\ell}\left(e^{j\omega T}\right)$ of the sequences
$c_{\ell}[n],1\leq\ell\leq K$, it was shown in \cite{location}
that the sequences $a_{\ell}\left[n\right],1\leq\ell\leq K$, which
define the signal $x\left(t\right)$, can be recovered from
$c_{\ell}[n]$ using an adequate multichannel filter bank.
Specifically, let $\mathbf{c}\left(e^{j\omega
T}\right),\mathbf{a}\left(e^{j\omega T}\right)$ denote the
length-$K$ column vectors whose $\ell$th elements are
$C_{\ell}\left(e^{j\omega T}\right),A_{\ell}\left(e^{j\omega
T}\right)$ respectively. Then, it can be shown that
\begin{equation}
\mathbf{c}\left(e^{j\omega
T}\right)=\mathbf{M}_{SG}\left(e^{j\omega
T}\right)\mathbf{a}\left(e^{j\omega T}\right),
\end{equation}
where $\mathbf{M}_{SG}\left(e^{j\omega T}\right)$ is a $K\times K$
matrix with $\ell k$th element \begin{equation}
\Phi_{S_{\ell}G_{k}}\left(e^{j\omega
T}\right)=\frac{1}{T}\sum_{m\in\mathbb{Z}}S_{\ell}^{*}\left(\omega-\frac{2\pi}{T}m\right)G_{k}\left(\omega-\frac{2\pi}{T}m\right).\end{equation}
Here $G_{k}\left(\omega\right)$ and $S_{\ell}\left(\omega\right)$
denote the Fourier transforms of $g_{k}\left(t\right)$ and
$s_{\ell}\left(t\right)$ respectively. If this matrix is
stably invertible a.e in $\omega$, then the sequences
$a_{\ell}\left[n\right]$ can be recovered from the samples using
the multichannel filter bank whose frequency response is given by
$\mathbf{M}_{SG}^{-1}\left(e^{j\omega T}\right)$, as depicted in
the right-hand side of Fig.~\ref{fig:SI}.

The proposed sampling scheme achieves an average sampling rate of
$K/T$ since there are $K$ sampling sequences each at rate $1/T$.
Intuitively, this approach requires one sample per degree of
freedom of the signal $x\left(t\right)$: for a signal of the form
\eqref{eq:x(t)}, under the assumption that the time delays are
known, or a signal of the form (\ref{eq:finitelygenerated}), in
each time period $T$ there are $K$ new parameters.

\subsection{Unknown delays}

We now address our original problem of designing a sampling scheme
for signals of the form \eqref{eq:x(t)} with unknown delays. We
propose a system similar to that used in the case of known delays,
comprised of parallel sampling channels. Since now there are more
degrees of freedom in the signal $x\left(t\right)$, intuitively we
will require at least the same number of channels as in the case
of known delays. Denoting the number of channels by $p$, this
implies that $p\geq K$. As we will see, under certain conditions
on the sampling filters and pulse $g(t)$, $p=2K$ sampling filters
are sufficient to guarantee perfect recovery of $x(t)$ from the
samples. We will also show that this is the minimal possible rate
achievable for all signals $x(t)$.

In each channel of our sampling system the signal $x\left(t\right)$ is
pre-filtered using the filter $s_{\ell}^{*}\left(-t\right)$ and
sampled uniformly at times $t=nT$ resulting in the samples
(\ref{eq:cn}). In the Fourier domain, we can write (\ref{eq:cn})
as
\begin{equation}
\label{eq:C(omega)} C_{\ell}\left(e^{j\omega T}\right)  =
\frac{1}{T}\sum_{m\in\mathbb{Z}}S_{\ell}^{*}\left(\omega-\frac{2\pi}{T}m\right)X\left(\omega-\frac{2\pi}{T}m\right).
\end{equation}
From the definition of $x(t)$, its Fourier transform can be
written as
\begin{eqnarray}
X\left(\omega\right) & = & \sum_{k=1}^{K}\sum_{n\in\mathbb{Z}}a_{k}\left[n\right]G\left(\omega\right)e^{-j\omega\left(t_{k}+nT\right)}\label{eq:X(omega)}\nonumber\\
 & = & \sum_{k=1}^{K}A_{k}\left(e^{j\omega T}\right)G\left(\omega\right)e^{-j\omega t_{k}}, \end{eqnarray}
where $A_{k}\left(e^{j\omega T}\right)$ denotes the DTFT of the
sequence $a_{k}\left[n\right]$, and $G\left(\omega\right)$ denotes
the Fourier transform of $g\left(t\right)$. Substituting
\eqref{eq:X(omega)} into \eqref{eq:C(omega)}, we have
\begin{eqnarray}
C_{\ell}\left(e^{j\omega T}\right)
 & = & \sum_{k=1}^{K}A_{k}\left(e^{j\omega T}\right)\frac{1}{T}\sum_{m\in\mathbb{Z}}S_{\ell}^{*}\left(\omega-\frac{2\pi}{T}m\right)\nonumber \\
 & & \cdot G\left(\omega-\frac{2\pi}{T}m\right)e^{-j\left(\omega-\frac{2\pi}{T}m\right)t_{k}}\nonumber \\
 & = & \sum_{k=1}^{K}A_{k}\left(e^{j\omega T}\right)e^{-j\omega t_{k}}\frac{1}{T}\sum_{m\in\mathbb{Z}}S_{\ell}^{*}\left(\omega-\frac{2\pi}{T}m\right)\nonumber \\
 & & \cdot G\left(\omega-\frac{2\pi}{T}m\right)e^{j\frac{2\pi}{T}mt_{k}},\label{eq:Cl(omega)} \end{eqnarray}
where the first equality is a result of the fact that
$A_{k}\left(e^{j\omega T}\right)$ is $2\pi/T$-periodic.

From now on, we will assume that $\omega \in \left[0,\frac{2\pi}{T}\right)$, and all
the expressions in the DTFT domain are $2\pi/T$ periodic.
Denoting by $\mathbf{c}\left(e^{j\omega T}\right)$ the length-$p$
column vector whose $\ell$th element is $C_{\ell}\left(e^{j\omega
T}\right)$ and by $\mathbf{a}\left(e^{j\omega T}\right)$ the
length-$K$ column vector whose $k$th element is
$A_{k}\left(e^{j\omega T}\right)$, we can write
\eqref{eq:Cl(omega)} in matrix form as
\begin{equation} \mathbf{c}\left(e^{j\omega
T}\right)=\mathbf{M}\left(e^{j\omega
T},\tau\right)\mathbf{D}\left(e^{j\omega
T},\tau\right)\mathbf{a}\left(e^{j\omega
T}\right).\label{eq:c(omega)}\end{equation}
Here
$\mathbf{M} \left(e^{j\omega T},\tau\right)$ is a $p\times K$ matrix with
 $\ell k$th element
\begin{eqnarray}
\mathbf{M}_{\ell k}\left(e^{j\omega T},\tau\right) & = & \frac{1}{T}\sum_{m\in\mathbb{Z}}S_{\ell}^{*}\left(\omega-\frac{2\pi}{T}m\right)\nonumber \\
& & \cdot G\left(\omega-\frac{2\pi}{T}m\right)
 e^{j\frac{2\pi}{T}mt_{k}}, \label{eq:Mlk(omega)} \end{eqnarray}
and $\mathbf{D}\left(e^{j\omega T},\tau\right)$ is a diagonal matrix with
$k$th diagonal element equal to $e^{-j\omega t_{k}}$. Defining the vector $\mathbf{b}\left(e^{j\omega
T}\right)$ as
\begin{equation}
\mathbf{b}\left(e^{j\omega
T}\right)\mathbf{\mathit{=}D}\left(e^{j\omega
T},\tau\right)\mathbf{a}\left(e^{j\omega
T}\right),\label{eq:b(omega)}\end{equation}
we can rewrite \eqref{eq:c(omega)} in the form \begin{equation}
\mathbf{c}\left(e^{j\omega T}\right)=\mathbf{M}\left(e^{j\omega
T},\tau\right)\mathbf{b}\left(e^{j\omega
T}\right).\label{eq:c(omega)3}\end{equation} Our problem can then
be reformulated as that of recovering $\mathbf{b}\left(e^{j\omega
T}\right)$ and the unknown delay set $\tau$ from the vectors
$\mathbf{c}\left(e^{j\omega T}\right)$, for all
$\omega\in\left[0,\frac{2\pi}{T}\right)$. Once these are known,
the vectors $\mathbf{a}\left(e^{j\omega T}\right)$ can be recovered
using the relation in \eqref{eq:b(omega)}.

To proceed, we focus our attention on sampling filters
$S_{\ell}(\omega)$ with finite support in the frequency domain,
contained in the frequency range \begin{equation}
\mathcal{F}=\left[\frac{2\pi}{T}\gamma,\frac{2\pi}{T}\left(p+\gamma\right)\right],\label{eq:workingband}\end{equation}
where $\gamma\in\mathbb{Z}$ is an index which determines the
working frequency band $\mathcal{F}$. This choice should be such
that it matches the frequency occupation of $g\left(t\right)$
(although $g(t)$ does not have to be bandlimited). This freedom
allows our sampling scheme to support both complex and real valued
signals. Under this choice of filters, each element
$\mathbf{M}\left(e^{j\omega T},\tau\right)$ of
(\ref{eq:Mlk(omega)}) can be expressed as\begin{equation}
\mathbf{M}_{\ell k}\left(e^{j\omega
T},\tau\right)=\sum_{m=1}^{p}\mathbf{W}_{\ell m}\left(e^{j\omega
T}\right)\mathbf{N}_{mk}\left(\tau\right),\label{eq:Mlk(omega)2}\end{equation}
where $\mathbf{W}\left(e^{j\omega T}\right)$ is a $p\times p$
matrix whose $\ell m$th element is given by\begin{eqnarray}
 \mathbf{W}_{\ell m}\left(e^{j\omega T}\right) & = & \frac{1}{T}S_{\ell}^{*}\left(\omega+\frac{2\pi}{T}\left(m-1+\gamma\right)\right) \nonumber \\
& & \cdot G\left(\omega+\frac{2\pi}{T}\left(m-1+\gamma\right)\right),
 \label{eq:Wlm(omega)}
  \end{eqnarray}
and $\mathbf{N}\left(\tau\right)$ is a $p\times K$ Vandermonde
matrix with $mk$th element\begin{equation}
\mathbf{N}_{mk}\left(\tau\right)=e^{-j\frac{2\pi}{T}\left(m-1+\gamma\right)t_{k}}.\label{eq:Nmk(omega)}\end{equation}
Substituting (\ref{eq:Mlk(omega)2}) into (\ref{eq:c(omega)3}),
\begin{equation}
\label{eq:Decomposition} \mathbf{c}(e^{j\omega
T})=\mathbf{W}\left(e^{j\omega
T}\right)\mathbf{N}\left(\tau\right)\mathbf{b}(e^{j\omega T}).
\end{equation}

If $\mathbf{W}\left(e^{j\omega T}\right)$ is stably invertible,
then we can define the modified measurement vector
$\mathbf{d}\left(e^{j\omega T}\right)$ as\begin{equation}
\mathbf{d}\left(e^{j\omega
T}\right)=\mathbf{W}^{-1}\left(e^{j\omega
T}\right)\mathbf{c}\left(e^{j\omega T}\right).\end{equation} This
vector satisfies \begin{equation}
\label{eq:doab}\mathbf{d}\left(e^{j\omega
T}\right)=\mathbf{N}\left(\tau\right)\mathbf{b}\left(e^{j\omega
T}\right).\end{equation} Since $\mathbf{N}\left(\tau\right)$ is
independent of $\omega$, from the linearity of the DTFT, we can
express \eqref{eq:doab} in the time domain as
\begin{equation}
\mathbf{d}\left[n\right]=\mathbf{N}\left(\tau\right)\mathbf{b}\left[n\right],\quad
n\in\mathbb{Z}.\label{eq:d[n]}\end{equation} The elements of the
vectors $\mathbf{d}\left[n\right]$ and $\mathbf{b}\left[n\right]$
are the discrete time sequences, obtained from the inverse DTFT of
the elements of the vectors $\mathbf{b}\left(e^{j\omega T}\right)$
and $\mathbf{d}\left(e^{j\omega T}\right)$ respectively.

\begin{figure}
\begin{center}
\includegraphics[scale=0.5]{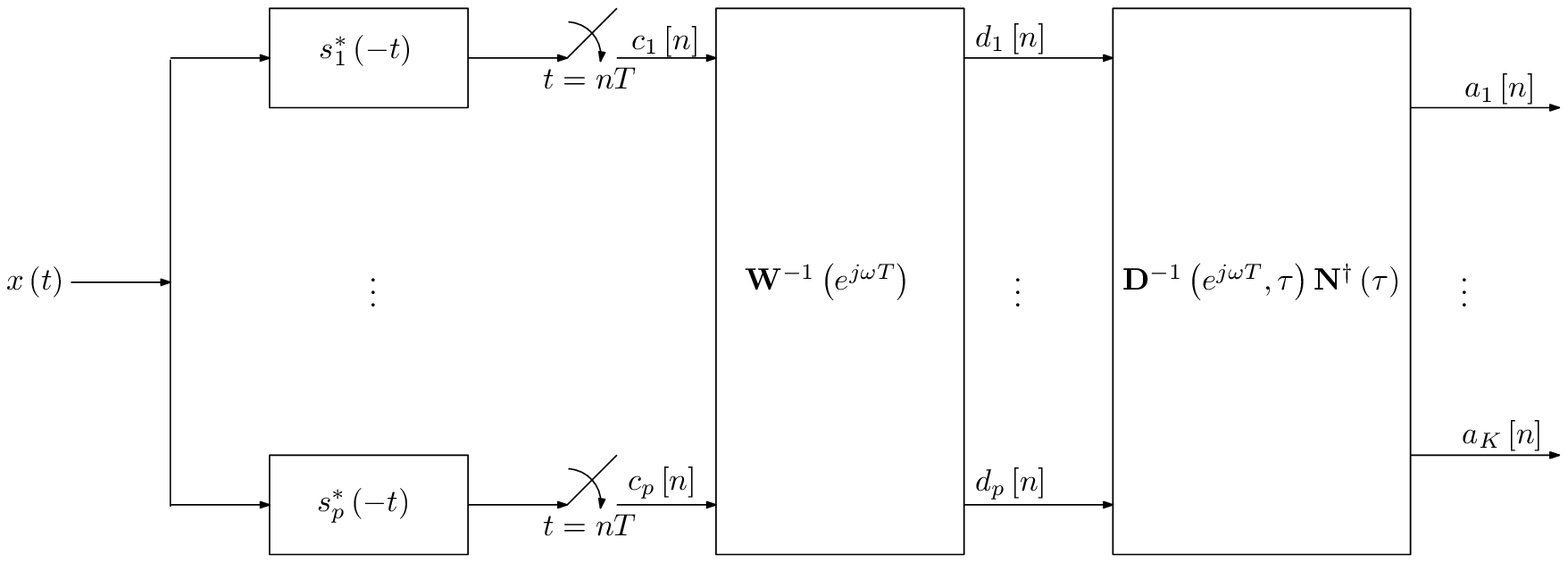}
\caption{Sampling and reconstruction scheme for the case of
unknown delays} \label{fig:figu}
\end{center}
\end{figure}

Equation \eqref{eq:doab} and its
equivalent time domain representation \eqref{eq:d[n]}, describe an
infinite set of measurement vectors, each obtained by the same
measurement matrix $\mathbf{N}\left(\tau\right)$, which depends on
the unknown delays $\tau$.  This problem is reminiscent of the
type of problems that arise in the field of DOA estimation \cite{krim1996tda},
as we discuss in the next section.
One class of efficient methods for DOA recovery, are known as subspace methods \cite{krim1996tda}.
These techniques have subsequently been applied to many other problems such
as spectral estimation \cite{stoica1997}, system identification \cite{moulines1995} and more.
Our approach is to rely on these methods
in order to first recover $\tau$ from the measurements. After $\tau$ is known, the vectors
$\mathbf{b}\left(e^{j\omega T}\right)$ and
$\mathbf{a}\left(e^{j\omega T}\right)$ can be found using linear
filtering relations by\begin{equation} \mathbf{b}\left(e^{j\omega
T}\right)=\mathbf{N^{\dagger}}\left(\tau\right)\mathbf{d}\left(e^{j\omega
T}\right).\end{equation} Since $\mathbf{N}\left(\tau\right)$ is a
Vandermonde matrix, its columns are linearly independent, and
consequently $\mathbf{N^{\dagger}\mathbf{N}}=\mathbf{I}_{K}$.
Using \eqref{eq:b(omega)},\begin{equation}
\mathbf{a}\left(e^{j\omega
T}\right)=\mathbf{D}^{-1}\left(e^{j\omega
T},\tau\right)\mathbf{N^{\dagger}}\left(\tau\right)\mathbf{d}\left(e^{j\omega
T}\right).\end{equation}
The resulting sampling and reconstruction scheme is depicted in
Fig.~\ref{fig:figu}.

Our last step, therefore, is to derive conditions on the filters
$s_{\ell}^{*}\left(-t\right)$ and the function $g\left(t\right)$
in order that the matrix $\mathbf{W}\left(e^{j\omega T}\right)$ is
stably invertible. To this end, we can decompose the matrix
$\mathbf{W}\left(e^{j\omega T}\right)$ as\begin{equation}
\mathbf{W}\left(e^{j\omega T}\right)=\mathbf{S}\left(e^{j\omega
T}\right)\mathbf{G}\left(e^{j\omega T}\right)\end{equation} where
$\mathbf{S}\left(e^{j\omega T}\right)$ is a $p\times p$ matrix
with $\ell m$th element \begin{equation} \mathbf{S}_{\ell
m}\left(e^{j\omega
T}\right)=\frac{1}{T}S_{\ell}^{*}\left(\omega+\frac{2\pi}{T}\left(m-1+\gamma\right)\right)\label{eq:Slm(omega)}\end{equation}
and $\mathbf{G}\left(e^{j\omega T}\right)$ is a $p\times p$
diagonal matrix whose $m$th diagonal element is given by
\begin{equation} \mathbf{G}_{mm}\left(e^{j\omega
T}\right)=G\left(\omega+\frac{2\pi}{T}\left(m-1+\gamma\right)\right).\label{eq:Gmm(omega)}\end{equation}
Each one of these matrices needs to be stably invertible.
Therefore, from \eqref{eq:Gmm(omega)} the condition that the
function $g\left(t\right)$ needs to satisfy is that
\begin{equation} 0<a\leq\left|G\left(\omega\right)\right|\leq
b<\infty\textrm{ a.e
}\omega\in\mathcal{F}.\label{eq:G(omega)}\end{equation} In
addition the filters $s_{\ell}^{*}\left(-t\right)$ should be
chosen in such a way that they form a stably invertible matrix
$\mathbf{S}\left(e^{j\omega T}\right)$. Examples of such filters
are given in the next subsection.

We note here that the conditions given above guarantee a stable
digital correction filter bank
$\mathbf{W}^{-1}\left(e^{j\omega T}\right)$, however generally it will be comprised of
infinite length digital filters. Practical implementation of these filters can be achieved
by truncating the impulse response.
The length of the resulting filters will affect the total delay of our proposed method, which will
generally be longer than that of the methods described in Section \ref{sec:Intro}, due to the additional
digital filtering stage.

We summarize the results so far in the following proposition.

\begin{prop}
\textit{\label{pro:d(omega)}}Let
$c_{\ell}\left[n\right]=\left\langle
x\left(t\right),s_{\ell}\left(t-nT\right)\right\rangle
,\;1\leq\ell\leq p$ be a set of $p$ sequences obtained by
filtering the signal $x\left(t\right)$ defined by \eqref{eq:x(t)}
with $p$ filters $s_{\ell}^{*}\left(-t\right)$ and sampling their
outputs at times $nT$. Let $S_{\ell}(\omega)$ be supported on
$\mathcal{F}=\left[\frac{2\pi}{T}\gamma,\frac{2\pi}{T}\left(p+\gamma\right)\right]$,
and let $\Omega=[0,\frac{2\pi}{T})$. If the function
$g\left(t\right)$ satisfies the condition in \eqref{eq:G(omega)}
and the matrix $\mathbf{S}\left(e^{j\omega T}\right)$, defined by
\eqref{eq:Slm(omega)}, is stably invertible a.e $\omega\in\Omega$,
then the delays $\tau$ and vector $\mathbf{b}\left(e^{j\omega
T}\right)$ can be found from the set of equations
\begin{equation}
\mathbf{d}\left(e^{j\omega
T}\right)=\mathbf{N}(\tau)\mathbf{b}\left(e^{j\omega T}\right),
\end{equation}
using subspace methods, described in the next section. Here
 $\mathbf{N}\left(\tau\right)$ is a $p\times
K$ Vandermonde matrix with $mk$th element
$e^{-j\frac{2\pi}{T}\left(m-1+\gamma\right)t_{k}}$, and
\begin{equation}
\mathbf{d}\left(e^{j\omega
T}\right)=\mathbf{W}^{-1}\left(e^{j\omega
T}\right)\mathbf{c}\left(e^{j\omega T}\right),
\end{equation}
with $\mathbf{W}\left(e^{j\omega T}\right)$ defined by
(\ref{eq:Wlm(omega)}). The sequences $a_{\ell}[n]$ can then be
recovered by
\begin{equation}
\mathbf{a}\left(e^{j\omega T}\right)=\mathbf{D}^{-1}\left(e^{j\omega T},
\tau\right)\mathbf{N}^\dagger(\tau) \mathbf{d}\left(e^{j\omega
T}\right),\quad \omega \in \Omega
\end{equation}
where $\mathbf{D}(e^{j\omega T,\tau})$ is a diagonal matrix with diagonal
elements $e^{-j\omega t_{k}}$.
\end{prop}
\subsection{Examples of filters}

We now provide some examples of filters $s_{\ell}\left(t\right)$
satisfying the required conditions.

\subsubsection{\label{sub:Complex-band-pass-filters}Complex bandpass filter-bank}

The first example is a set of complex bandpass filters. We assume
that the working band is
$\mathcal{F}=\left[0,\frac{2\pi}{T}p\right]$ ($\gamma=0)$, and the
function $g\left(t\right)$ satisfies \eqref{eq:G(omega)} on that
frequency range. We choose the filters
$s_{\ell}^{*}\left(-t\right)$ as ideal bandpass filters, covering
consecutive frequency bands:\begin{equation}
S_{\ell}\left(\omega\right)=\begin{cases}
T, & \omega\in\left[\left(\ell-1\right)\frac{2\pi}{T},\ell\frac{2\pi}{T}\right]\\
0, & \textrm{otherwise.}\end{cases}\end{equation} The resulting
matrix $\mathbf{S}\left(e^{j\omega T}\right)$ is diagonal, and
 stably invertible.
This example generalizes to any valid working band, given by
\eqref{eq:workingband}, by shifting the frequency response of the
filters.

\begin{figure}
\begin{center}
\subfigure[]{\includegraphics[scale=0.3]{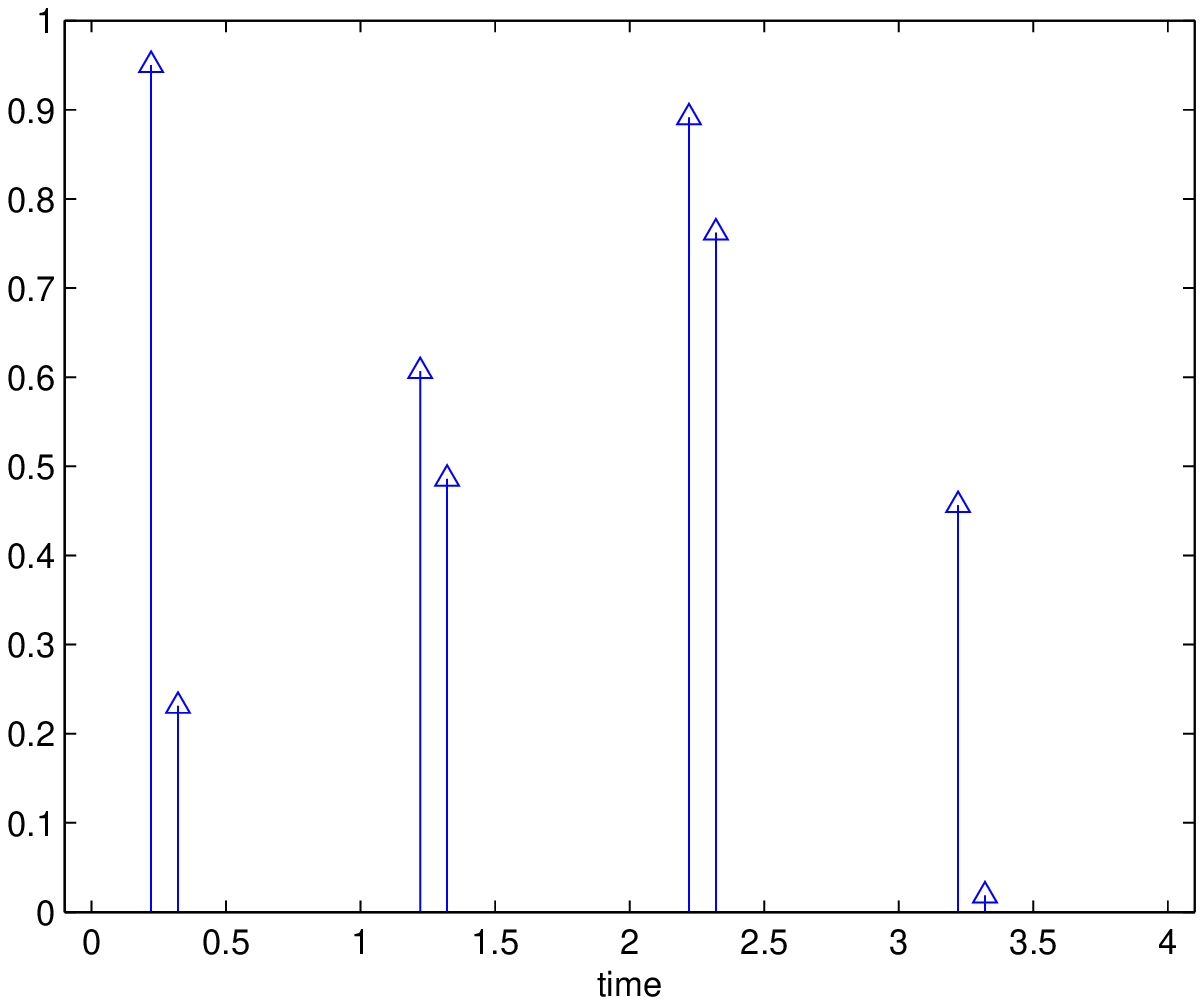}}\subfigure[]{\includegraphics[scale=0.3]{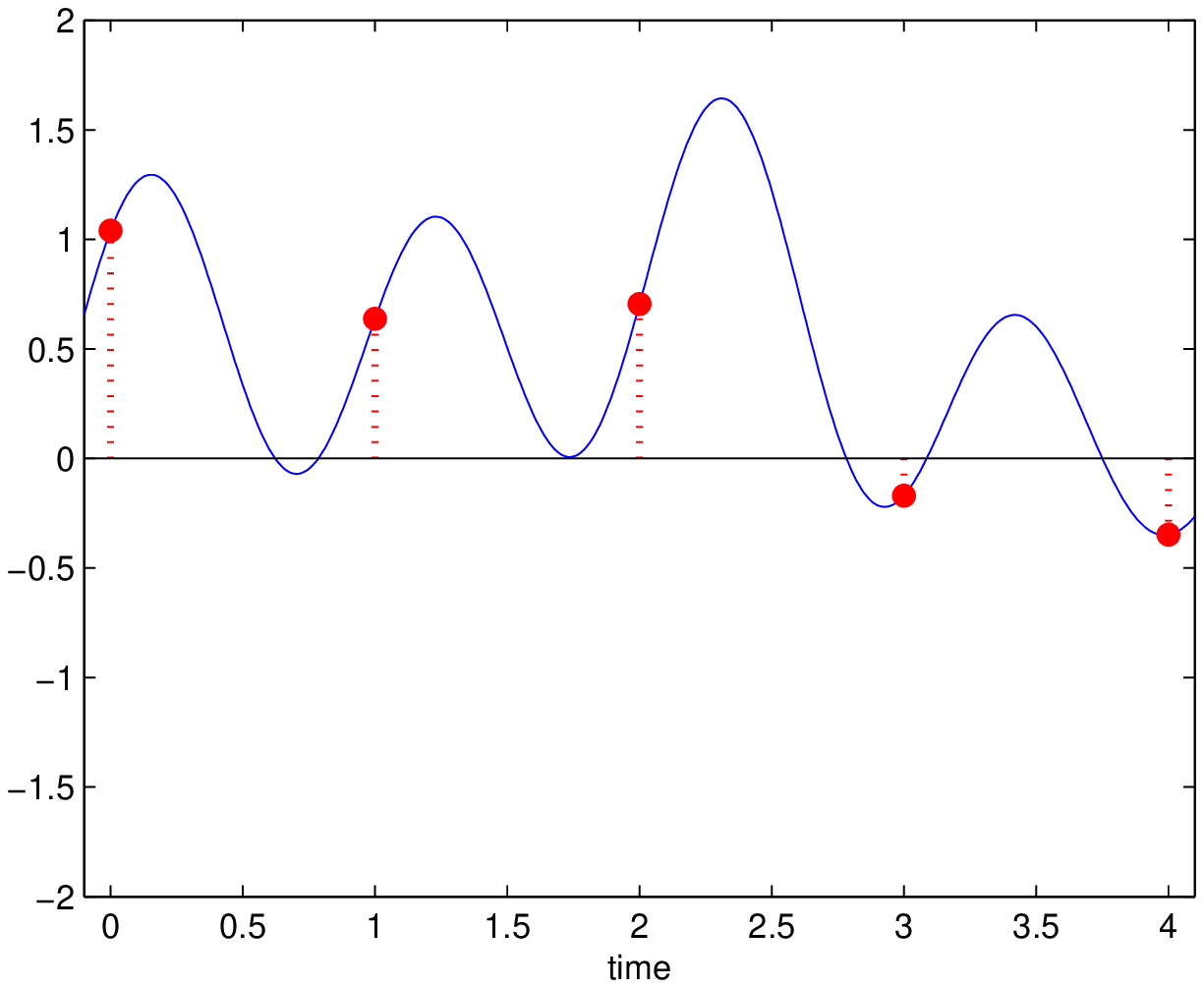}}

\subfigure[]{\includegraphics[scale=0.3]{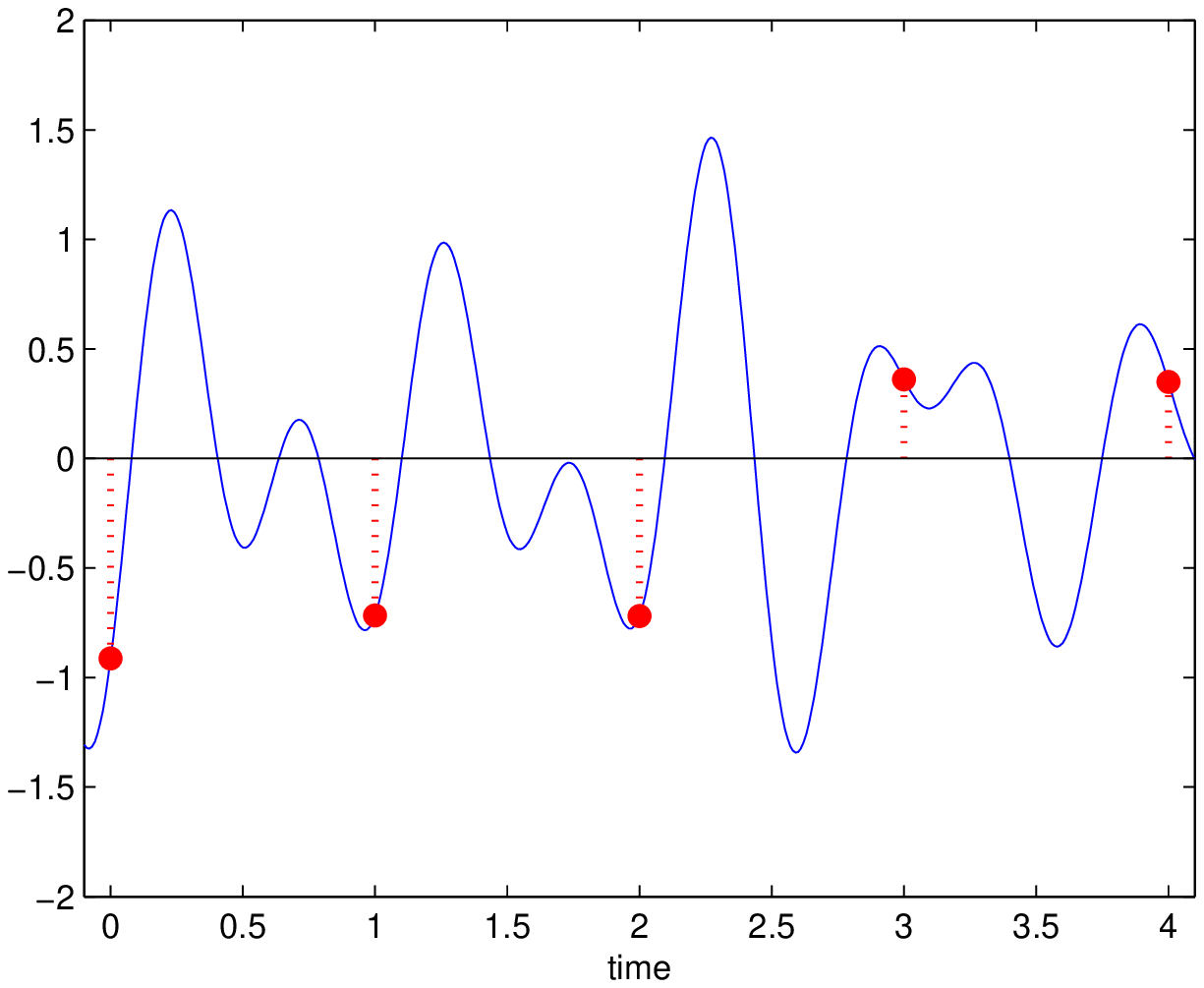}}\subfigure[]{\includegraphics[scale=0.3]{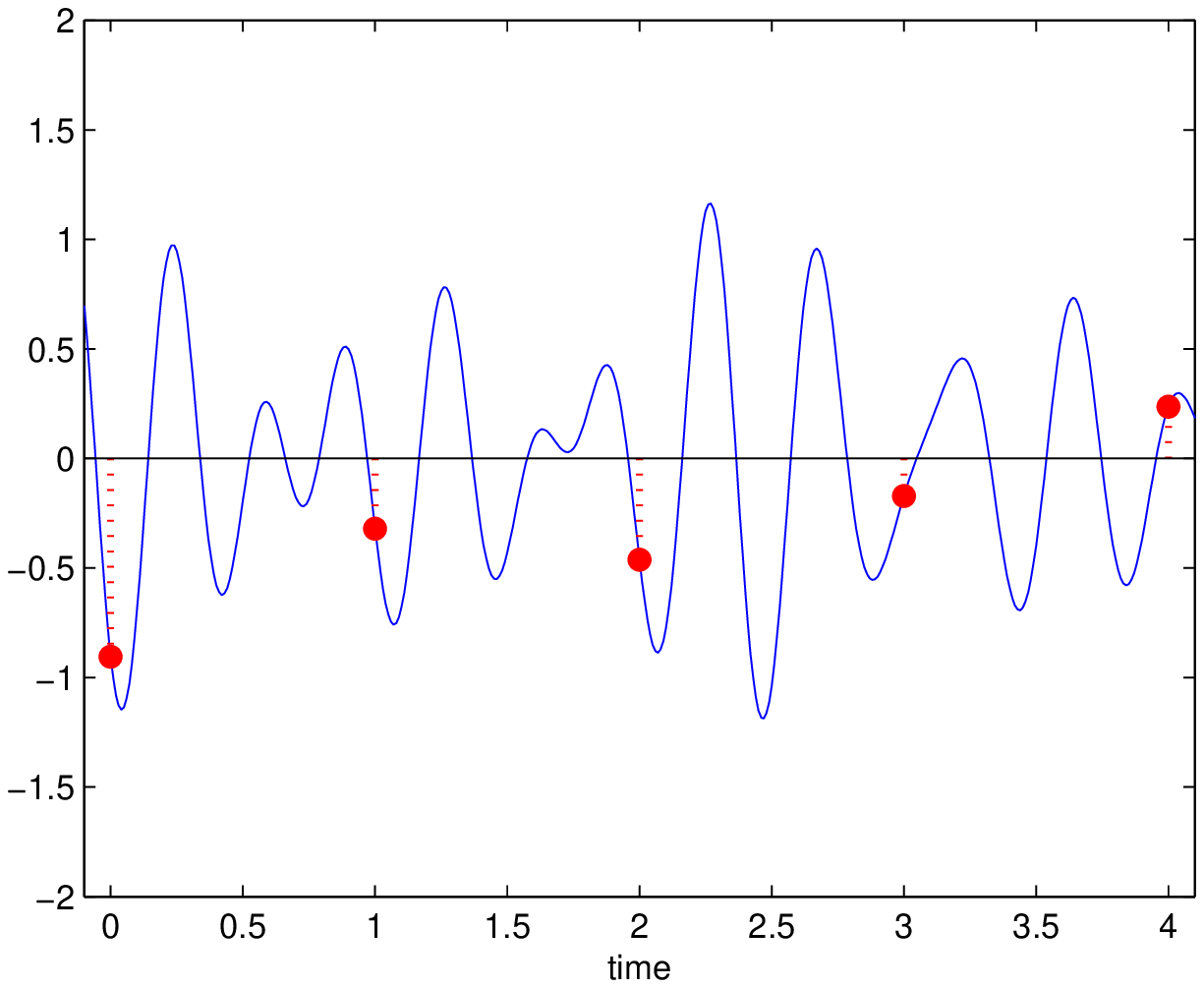}}

\caption{\label{fig:sampling}Stream of diracs. (a) $K=2$ diracs
per period $T=1$. (b)-(d) The outputs of the first three sampling
channels. The dashed lines denote the sampling instants.}
\end{center}
\end{figure}

We now provide an example demonstrating the importance of the
sampling filter.
\begin{example}
We consider the case where $g(t)=\delta(t)$ and there are $K=2$
diracs per period of $T=1$, as illustrated in
Fig.~\ref{fig:sampling}(a). The sampling scheme described above,
consisting of a complex bandpass filter-bank, is used. In
Figs.~\ref{fig:sampling}(b)--(d), we show the outputs of the first
$3$ sampling channels. This example demonstrates the need for the
sampling filters when sampling short-length pulses at a low
sampling rate. The sampling kernels have the effect of smoothing
the short pulses (diracs in this example). Consequently, even when
the sampling rate is low, the samples contain information about
the signal. In contrast, if we were to sample the signal in
Fig.~\ref{fig:sampling}(a) directly at a low rate, then we would
often obtain only zero samples which contain no information about
the signal.
\end{example}

\subsubsection{Delayed channels}

In this example we assume $p$ is even and define the working band
as \begin{equation}
\mathcal{F}=\left[-\frac{\pi}{T}p,\frac{\pi}{T}p\right]\end{equation}
($\gamma=-p/2)$. We also assume that $g\left(t\right)$ satisfies
\eqref{eq:G(omega)}. We choose the $\ell$th filter as a delay of
$\Delta_{\ell}\in\left[0,T\right)$ followed by an ideal low pass
filter. Thus, \begin{equation}
S_{\ell}\left(\omega\right)=\begin{cases}
Te^{-j\omega\Delta_{\ell}}, & \omega\in\mathcal{F}\\
0, & \textrm{otherwise.}\end{cases}\end{equation} With this choice
of filters, the $\ell m$th element of the matrix
$\mathbf{S}\left(e^{j\omega T}\right)$ defined in
\eqref{eq:Slm(omega)} is given by\begin{eqnarray}
\mathbf{S}_{\ell m}\left(e^{j\omega T}\right) & = & e^{j\left(\omega+\frac{2\pi}{T}\left(m-1-p/2\right)\right)\Delta_{\ell}}\nonumber\\
 & = & e^{j\left(\omega-\frac{2\pi}{T}\left(p/2\right)\right)\Delta_{\ell}}e^{j\frac{2\pi}{T}\left(m-1\right)\Delta_{\ell}}.\end{eqnarray}
The matrix $\mathbf{S}\left(e^{j\omega T}\right)$ can be expressed
as\begin{equation}
\mathbf{S}\left(e^{j\omega T}\right)=\mathbf{\Phi}\left(e^{j\omega T}\right)\mathbf{F},\label{eq:S_delay}\end{equation}
where $\mathbf{\Phi}\left(e^{j\omega T}\right)$ is a diagonal $p\times p$
matrix whose $\ell$th diagonal element is $e^{j\left(\omega-\frac{2\pi}{T}\left(p/2\right)\right)\Delta_{\ell}}$,
and $\mathbf{F}$ is a Vandermonde matrix whose $\ell m$th element
is given by $e^{j\frac{2\pi}{T}\left(m-1\right)\Delta_{\ell}}$. From \eqref{eq:S_delay}
it can be seen that $\mathbf{S}\left(e^{j\omega T}\right)$ is invertible
for all $\omega\in\Omega$, when the delays in each channel $\Delta_{\ell}$
are distinct.

One special case of this choice of sampling filters is when the
delays are uniformly spaced, i.e
$\Delta_{\ell}=\left(\ell-1\right)T/p$. In this case our sampling
scheme can be implemented by an ideal low pass filter with cutoff
$\pi p/T$, followed by a uniform sampler at a rate of $p/T$.

\section{\label{sec:Unknown-Delays-Recovery}Recovery of the Unknown Delays}

We have seen in the previous section that perfect reconstruction
of a signal $x\left(t\right)$ of the form \eqref{eq:x(t)}, is
equivalent to that of recovering the delays $\tau$ from the
modified measurements of \eqref{eq:d[n]}. As we now show, this
problem is similar to that arising in DOA estimation.

\subsection{Relation to direction of arrival estimation}

In DOA estimation \cite{1143830,32276,Johnson1993,krim1996tda},
$K$ narrow band sources impinge on an array, composed of $p$
sensors, from distinct DOAs. The goal is to estimate the DOAs of
the sources from a set of $M$ measurements, obtained from the
sensors outputs at distinct time instants.

The DOA estimation problem can be formulated using the following
measurement model \begin{equation}
\boldsymbol{X}=\boldsymbol{A}\left(\Theta\right)\mathbf{S}\label{eq:X}\end{equation}
where $\boldsymbol{X}$ is a $p\times M$ matrix, composed of the
measurements in its columns, $\mathbf{S}$ is a $K\times M$ matrix
consisting of the sources signals in its columns and
$\boldsymbol{A}\left(\Theta\right)$ is a $p\times K$ matrix which
depends on the set of unknown DOAs $\Theta$. The structure of
$\boldsymbol{A}\left(\Theta\right)$ is such that its $k$th column,
denoted $\mathbf{a}\left(\theta_{k}\right)$, depends only on the
DOA $\theta_{k}$ of the $k$th source. The vector
$\mathbf{a}\left(\theta_{k}\right)$ is referred to as the steering
vector of the array toward direction $\theta_{k}$. The set
containing all possible steering vectors, i.e, $\left\{
\mathbf{a}\left(\theta\right),\theta\in\left[0,2\pi\right)\right\}
$ is referred as the array manifold. Given $\boldsymbol{X}$, the
problem is to recover the DOAs $\theta_k$, and the sources
$\boldsymbol{S}$.

The set of equations in \eqref{eq:d[n]} has the same form as
(\ref{eq:X}). The $k$th column of the matrix
$\mathbf{N}\left(\tau\right)$ depends only on the $k$th unknown
delay $t_k$, and can be described by the vector
$\mathbf{n}\left(t_{k}\right)$, where
\begin{equation}
\mathbf{n}\left(t\right)=\left[\begin{array}{cccc}
e^{-j\frac{2\pi}{T}\gamma t} &
e^{-j\frac{2\pi}{T}\left(1+\gamma\right)t} & \cdots &
e^{-j\frac{2\pi}{T}\left(p-1+\gamma\right)t}\end{array}\right]^{T}.\label{eq:n(t)}\end{equation}
The array manifold in our setting is the set of vectors $\{
\mathbf{n}(t),t\in[0,T)\}$. Therefore, we can adopt DOA methods to
estimate the unknown delays. The only difference between the two
problems is that in our setting, we have infinitely many
measurement vectors, in contrast to the DOA problem in which
$\boldsymbol{X}$ has finitely many columns. This will require
several adjustments, which we will detail in the ensuing
subsections.

Two prominent methods used for DOA estimation are MUSIC (MUltiple
Signal Classification) \cite{1143830} and ESPRIT (Estimation of
Signal Parameters via Rotational Invariance Techniques)
\cite{32276}. These algorithms belong to a class of techniques,
known as subspace methods, which are based on separating the space
containing the measurements into two subspaces, the signal and
noise subspaces. Estimating the unknown set of parameters using
MUSIC involves a continuous one-dimensional search over the
parameter range. This procedure can be costly from a computational
point of view. The ESPRIT approach can estimate the unknown set of
parameters more efficiently, by imposing the additional
requirement that the measurement matrix is rotationally invariant.
We describe this property in subsection
\ref{sub:Recovering-the-unknown} and show that in our case
$\mathbf{N}\left(\tau\right)$ satisfies this condition, and
therefore we use the ESPRIT approach.

We note that although the MUSIC and ESPRIT methods were originally developed
as DOA estimators, these approaches and other subspace methods have been used successfully in
other fields.

\subsection{Sufficient conditions for perfect recovery}

We now rely on results obtained in the context of DOA estimation
in order to develop sufficient conditions for a unique solution to
\eqref{eq:d[n]}. Such a solution consists of the infinite set of
vectors $\mathbf{b}\left[n\right], n\in\mathbb{Z}$ and the unknown
delays $\tau.$

Conditions for a unique solution $\left(\Theta,\mathbf{S}\right)$
for \eqref{eq:X} where derived in \cite{32277}.
 Since \cite{32277} deals with a finite number of
measurements, we have to extend the results to our case, which
consists of an infinite number of measurements. The analysis in
\cite{32277} requires a preliminary condition that any subset of
$p$ distinct steering vectors from the array manifold is linearly
independent.  In our case this condition translates into the
requirement that any set of $p$ vectors
$\mathbf{n}\left(t_{i}\right),1\leq i\leq p$ associated with
distinct delays $t_{i}\in\left[0,T\right),1\leq i\leq p$ are
linearly independent. From \eqref{eq:n(t)}, any such set forms a
$p\times p$ Vandermonde matrix, and are therefore linearly
independent. Therefore, this condition automatically holds in our
problem without forcing any additional constraints.

To derive sufficient conditions for a unique solution of the set
of infinite equations \eqref{eq:d[n]} we introduce some notation.
We define the measurement set $\mathbf{d}\left[\Lambda\right]$ as
the set containing all measurement vectors
$\mathbf{d}\left[\Lambda\right]=\left\{
\mathbf{d}\left[n\right],n\in\mathbb{Z}\right\} $. Similarly, we
define the unknown vector set $\mathbf{b}\left[\Lambda\right]$ as
$\mathbf{b}\left[\Lambda\right]=\left\{
\mathbf{b}\left[n\right],n\in\mathbb{Z}\right\} $. We may then
rewrite \eqref{eq:d[n]} as
\begin{equation}
\mathbf{d}\left[\Lambda\right]=\mathbf{N}\left(\tau\right)\mathbf{b}\left[\Lambda\right].\label{eq:d[Lambda]}\end{equation}
The following proposition provides sufficient conditions for a
unique solution to \eqref{eq:d[Lambda]}.

\begin{prop}
\textit{\label{pro:unique}}If
$\left(\bar{\tau},\bar{\mathbf{b}}\left[\Lambda\right]\right)$ is
a solution to \eqref{eq:d[Lambda]},\begin{equation}
p>2K-\textrm{dim}\left(\textrm{span}\left(\bar{\mathbf{b}}\left[\Lambda\right]\right)\right)\label{eq:p>2K-dim}\end{equation}
and \begin{equation}
\textrm{dim}\left(\textrm{span}\left(\bar{\mathbf{b}}\left[\Lambda\right]\right)\right)\geq1,\label{eq:dimspan(b)}\end{equation}
then
$\left(\bar{\tau},\bar{\mathbf{b}}\left[\Lambda\right]\right)$ is
the unique solution of \eqref{eq:d[Lambda]}.
\end{prop}
The notation
$\textrm{span}\left(\bar{\mathbf{b}}\left[\Lambda\right]\right)$
is used for the minimal dimension subspace containing the unknown
vector set $\bar{\mathbf{b}}\left[\Lambda\right]$. The condition
\eqref{eq:dimspan(b)} is needed to avoid the case where
$\bar{\mathbf{b}}\left[\Lambda\right]=0.$ In this case clearly the
set $\tau$ can not recovered uniquely.

\begin{proof}
We denote
$r=\textrm{dim}\left(\textrm{span}\left(\bar{\mathbf{b}}\left[\Lambda\right]\right)\right)$.
From \eqref{eq:dimspan(b)} $r\geq1$. Therefore, there exist a
finite subset $\tilde{\Lambda}=\left\{ n_{i}\right\}
_{i=1}^{r}\subset\Lambda$, such that the vector set
$\bar{\mathbf{b}}\tilde{\left[\Lambda\right]}$ spans an
$r$-dimensional subspace: \begin{equation}
\textrm{dim}\left(\textrm{span}\left(\bar{\mathbf{b}}\tilde{\left[\Lambda\right]}\right)\right)=r.\end{equation}
By defining the matrices $\mathbf{B}$ and $\mathbf{D}$ as the
matrices consisting of the vector sets
$\bar{\mathbf{b}}\tilde{\left[\Lambda\right]}$ and
$\bar{\mathbf{d}}\tilde{\left[\Lambda\right]}$, we can write
\begin{equation}
\mathbf{D}=\mathbf{N}\left(\bar{\tau}\right)\mathbf{B}.\label{eq:D}\end{equation}
Clearly, from its construction, the rank of the matrix $\mathbf{B}$
is $r$. From \eqref{eq:p>2K-dim}, \begin{equation}
p>2K-r.\label{eq:p>2K-r}\end{equation}
According to Theorem 1 in \cite{32277}, the solution $\left(\bar{\tau},\mathbf{B}\right)$
is the unique solution to \eqref{eq:D} under the condition \eqref{eq:p>2K-r}.

Since the set of unknown delays $\bar{\tau}$ is the unique solution
to the finite set of equations \eqref{eq:D}, it is also a unique
solution to the infinite set of equations \eqref{eq:d[Lambda]}. Once
$\bar{\tau}$ is uniquely determined, the matrix $\mathbf{N}\left(\bar{\tau}\right)$
is known. Since every vector of the vector set $\bar{\mathbf{b}}\tilde{\left[\Lambda\right]}$
is contained in $\mathbb{C}^{K}$, \begin{equation}
\textrm{dim}\left(\textrm{span}\left(\bar{\mathbf{b}}\tilde{\left[\Lambda\right]}\right)\right)\leq K.\end{equation}
Therefore, according to \eqref{eq:p>2K-dim} $p>K$. The matrix $\mathbf{N}\left(\bar{\tau}\right)$
is a $p\times K$ Vandermonde matrix which consist of $K$ linearly
independent vectors. Therefore, for every $n\in\Lambda$, if $\bar{\mathbf{b}}\left[n\right]$
is a solution to \begin{equation}
\mathbf{d}\left[n\right]=\mathbf{N}\left(\bar{\tau}\right)\mathbf{b}\left[n\right]\end{equation}
then it is the unique solution.
\end{proof}
Proposition \ref{pro:unique} suggests that a unique solution to
the set of equations \eqref{eq:d[n]} is guaranteed, under proper
selection of the number of sampling channels $p$. This parameter,
in turn, determines the average sampling rate of our sampling
scheme, which is given by $p/T$. The condition \eqref{eq:p>2K-dim}
depends on the value of
$\textrm{dim}\left(\textrm{span}\left(\mathbf{b}\left[\Lambda\right]\right)\right)$,
which is generally not known in advance. According to our
assumption
$\textrm{dim}(\textrm{span}(\bar{\mathbf{b}}\tilde{[\Lambda]}))\geq1$,
therefore in order to satisfy the uniqueness condition
\eqref{eq:p>2K-dim} for every signal of the form \eqref{eq:x(t)},
we must have $p>2K-1$ sampling channels or a minimal sampling rate
of $2K/T$. Comparing this result to the minimal sampling rate in
the case when the delays are known in advance, there is a penalty
of $2$ in the minimal rate.

In Section~\ref{sub:union} we show that our signal
model, described in \eqref{eq:x(t)}, can be considered as part of
a more general framework of signals that lie in a union of SI subspaces \cite{4483755}.
It was shown in  \cite{4483755} that the theoretical minimum sampling rate
required for perfect recovery of such a signal from its samples is
$2K/T$. Therefore, according to the results of
Proposition \ref{pro:unique}, our sampling scheme can achieve the
minimal sampling rate required for signals of the form
\eqref{eq:x(t)}.

The minimal sampling rate of $2K/T$, which is achieved by our scheme,
does not depend on the bandwidth of the pulse $g\left(t\right)$,
but only on the number of propagation paths $K$.
In applications where the number of propagation paths is relatively small, or the
bandwidth of the transmitted pulse is high, our approach can provide
a sampling rate lower than the Nyquist rate. More precisely, when $2K/T<W$, where $W$
is the bandwidth of transmitted pulse, our method can reduce the sampling rate relatively to the
Nyquist rate. For example, the setup in \cite{uwb_char}, used for characterization of ultra-wide band (UWB) wireless
indoor channels, consists of pulses with bandwidth of $W=1$GHz transmitted at a rate of $1/T=2$MHz. Under the assumption that there are $32$ significant multipath components, our method can reduce the sampling rate down to
$128$MHz compared with the $2$GHz Nyquist rate.

Beside the theoretical interest, sampling rate reduction is also important for implementation considerations. For practical ADCs, which perform the sampling process, there is a trade-off between sampling rate and resolution \cite{ADC}. Therefore, reducing the sampling rate allows the use of more precise ADCs, which can improve the time delay estimation. The power consumption of an ADC can also be reduced by lowering the sampling rate \cite{ADC}.
In addition, a lower rate  leads to more efficient digital processing hardware, since a smaller number of samples has to be processed. This also allows performing the digital processing operations in real time.

\subsection{\label{sub:Recovering-the-unknown}Recovering the unknown delays}

According to Proposition \ref{pro:unique}, in order be able to
perfectly reconstruct every signal of the form \eqref{eq:x(t)},
our sampling scheme must have $p\geq2K$ sampling channels. We
assume throughout that this condition holds.

We now describe an algorithm for the recovery of the unknown
delays from the measurement set $\mathbf{d}\left[\Lambda\right]$,
which is based on the ESPRIT \cite{32276} algorithm. One of the
conditions needed in order to use the ESPRIT method is that the
correlation matrix
\begin{equation}
\mathbf{R}_{bb}=\sum_{n\in\mathbb{Z}}\mathbf{b}\left[n\right]\mathbf{b}^{H}\left[n\right],\label{eq:Rbb}\end{equation}
is positive definite. In order to relate this condition to our
problem, we state the following proposition from \cite{4553693}.

\begin{prop}
\label{pro:Rbb}If the sum \eqref{eq:Rbb} exists, then every matrix
$\mathbf{V}$ satisfying $\mathbf{R}_{bb}=\mathbf{V}\mathbf{V}^{H}$
has column span equal to $\textrm{span}\left(\mathbf{b}\left[\Lambda\right]\right)$.

An immediate corollary from Proposition \ref{pro:Rbb} is that
$\mathbf{R}_{bb}\succ0$ is equivalent to the condition
$\textrm{dim}\left(\textrm{span}\left(\mathbf{b}\left[\Lambda\right]\right)\right)=K$.
In this case, which we refer to as the \textit{uncorrelated case},
we can apply the ESPRIT algorithm on the measurement set
$\mathbf{d}\left[n\right]$ in order to recover the unknown delays.
The case
$\textrm{dim}\left(\textrm{span}\left(\mathbf{b}\left[\Lambda\right]\right)\right)<K$,
will be referred to as the\textit{ correlated case}. In this
setting the condition $\mathbf{R}_{bb}\succ0$ doest not hold, and
the ESPRIT algorithm cannot applied directly. Instead, we will use
an additional stage originally proposed in \cite{title,1164649}.
\end{prop}

Note, that
\begin{eqnarray}
\mathbf{R}_{dd} & = & \sum_{n\in\mathbb{Z}}\mathbf{d}\left[n\right]\mathbf{d}^{H}\left[n\right]\label{eq:Rdd}\nonumber \\
 & = & \mathbf{N}\left(\tau\right)\left(\sum_{n\in\mathbb{Z}}\mathbf{b}\left[n\right]\mathbf{b}^{H}\left[n\right]\right)\mathbf{N}^{H}\left(\tau\right)\nonumber \\
 & = & \mathbf{N}\left(\tau\right)\mathbf{R}_{bb}\mathbf{N}^{H}\left(\tau\right).\end{eqnarray}
Since for any set of delays $t_k$, the matrix
$\mathbf{N}\left(\tau\right)$ has full column-rank, the ranks of
the matrices $\mathbf{R}_{dd}$ and $\mathbf{R}_{bb}$ are equal.
Therefore, the decision whether we are in the uncorrelated or
correlated case can made directly from the given measurements by
forming the matrix $\mathbf{R}_{dd}$.

\subsubsection{Uncorrelated Case}

From \eqref{eq:Rdd}, under the assumption that the matrix
$\mathbf{R}_{bb}$ is positive definite, it can be shown that the
rank of the matrix $\mathbf{R}_{dd}$ is $K$. Moreover, the
matrices $\mathbf{R}_{dd}$ and $\mathbf{N}\left(\tau\right)$ have
the same column span which is referred as the signal subspace. By
performing a singular value decomposition (SVD) of the matrix
$\mathbf{R}_{dd}$, we can obtain $K$ vectors, which span the
signal subspace, by taking the $K$ left singular vectors
associated to the non-zero singular values of $\mathbf{R}_{dd}$.
We define the $p\times K$ matrix $\mathbf{E}_{s}$ as the matrix
containing those vectors in its columns.

Now, we will the exploit the special structure of the Vandermonde
matrix. We denote the matrix
$\mathbf{N}_{\downarrow}\left(\tau\right)$ as the sub matrix
extracted from $\mathbf{N}\left(\tau\right)$ by deleting its last
row. In the same way we define
$\mathbf{N}_{\uparrow}\left(\tau\right)$ as the sub matrix
extracted from $\mathbf{N}\left(\tau\right)$ by deleting its first
row. The Vandermonde matrix $\mathbf{N}\left(\tau\right)$
satisfies the following rotational invariance
property:\begin{equation}
\mathbf{N}_{\uparrow}\left(\tau\right)=\mathbf{N}_{\downarrow}\left(\tau\right)\mathbf{R}\left(\tau\right)\label{eq:rotate}\end{equation}
where $\mathbf{R}\left(\tau\right)$ is a diagonal $K\times K$
matrix, whose $k$th diagonal element is given by $
\mathbf{R}_{kk}\left(\tau\right)=e^{-j2 \pi t_{k}/T}$. Since the
matrices $\mathbf{N}\left(\tau\right)$ and $\mathbf{E}_{s}$ have
the same column span, there exists an invertible $K\times K$
matrix $\mathbf{T}$ such that\begin{equation}
\mathbf{N}\left(\tau\right)=\mathbf{E}_{s}\mathbf{T}.\label{eq:N(tau)}\end{equation}
By deleting the last row in \eqref{eq:N(tau)} we
get\begin{equation}
\mathbf{N}_{\downarrow}\left(\tau\right)=\mathbf{E}_{s\downarrow}\mathbf{T}.\label{eq:Ndown}\end{equation}
Similarly, deleting the first row in \eqref{eq:N(tau)} and using
the rotational invariance property \eqref{eq:rotate}, we have
\begin{equation}
\mathbf{N}_{\downarrow}\left(\tau\right)\mathbf{R}\left(\tau\right)=\mathbf{E}_{s\uparrow}\mathbf{T}.\label{eq:Nup}\end{equation}
Combining \eqref{eq:Ndown} and \eqref{eq:Nup} leads to the
following relation between the matrices
$\mathbf{\mathbf{E}_{s\uparrow}}$ and $\mathbf{E}_{s\downarrow}$:
\begin{equation}
\mathbf{\mathbf{E}_{s\uparrow}}=\mathbf{E}_{s\downarrow}\mathbf{T}\mathbf{R}\left(\tau\right)\mathbf{T}^{-1}.\label{eq:Esup}\end{equation}
The matrix $\mathbf{\mathbf{E}_{s\uparrow}}$ is a
$\left(p-1\right)\times K$ ($p>K)$ matrix with full column rank.
Therefore
$\mathbf{E}_{s\downarrow}^{\dagger}\mathbf{\mathbf{E}_{s\uparrow}}=\mathbf{I}_{K}$.
Using \eqref{eq:Esup} we define the following $K\times K$ matrix
$\mathbf{\Phi}$ as\begin{equation}
\mathbf{\Phi=\mathbf{E}_{s\downarrow}^{\dagger}}\mathbf{\mathbf{E}_{s\uparrow}}=\mathbf{T}\mathbf{R}\left(\tau\right)\mathbf{T}^{-1}.\label{eq:Phi}\end{equation}
From \eqref{eq:Phi} it is clear that the diagonal matrix
$\mathbf{R}\left(\tau\right)$ can be obtained from the matrix
$\mathbf{\Phi}$ by performing an eigenvalue decomposition. Once
the matrix $\mathbf{R}\left(\tau\right)$ is known, the unknown
delays can be retrieved from its diagonal elements
as\begin{equation}
t_{k}=-\frac{T}{2\pi}\textrm{arg}\left(\mathbf{R}_{kk}\left(\tau\right)\right).\end{equation}

In summary, our algorithm consist of the following steps:

\begin{enumerate}
\item Construct the correlation matrix
$\mathbf{R}_{dd}=\sum_{n\in\mathbb{Z}}\boldsymbol{d}\left[n\right]\boldsymbol{d}^{H}\left[n\right]$.
\item Perform an SVD decomposition of $\mathbf{R}_{dd}$ and
construct the matrix $\mathbf{E}_{s}$ consisting of the $K$
singular vectors associated with the non-zero singular values in
its columns. \item Compute the matrix
$\mathbf{\Phi=\mathbf{E}_{s\downarrow}^{\dagger}}\mathbf{\mathbf{E}_{s\uparrow}}$.
\item Compute the eigenvalues of $\mathbf{\Phi}$,
$\lambda_{i},i=1,2,\ldots,K$.
\item Retrieve the unknown delays by $t_{i}=-\frac{T}{2\pi}\textrm{arg}\left(\lambda_{i}\right)$.\\

\end{enumerate}

\subsubsection{Correlated Case}

When the condition $\mathbf{R}_{bb}\succ0$ is not satisfied the
ESPRIT algorithm cannot be applied directly on the vector set
$\mathbf{d}\left[\Lambda\right]$. In this case the rank of
$\mathbf{R}_{dd}$ is smaller than $K$, and therefore its column
span is no longer equal to the entire signal subspace. To
accommodate this setting, we perform an additional stage before
applying the ESPRIT method, based on the spatial smoothing
technique proposed in \cite{1164649,title}.

To proceed, we  define $M=p-K$ length-$\left(K+1\right)$ sub
vectors
\begin{equation}
\boldsymbol{d}_{i}\left[n\right]=\left[\begin{array}{cccc}
d_{i}\left[n\right] & d_{i+1}\left[n\right] & \ldots &
d_{i+K}\left[n\right]\end{array}\right]^{T}.\end{equation}
We define the smoothed correlation matrix $\overline{\mathbf{R}}_{dd}$ as
\begin{equation}
\mathbf{\overline{R}}_{dd}=\frac{1}{M}\sum_{i=1}^{M}\sum_{n\in\mathbb{Z}}\boldsymbol{d}_{i}\left[n\right]\boldsymbol{d}_{i}^{H}\left[n\right].\end{equation}

Under our assumptions $p\geq2K$, therefore $M\geq K$. According to
\cite{1164649}, when $M\geq K$ the rank of the smoothed correlation matrix
is $K$ regardless of the rank of the matrix $\mathbf{R}_{bb}$.
We will refer now to column rank of $\overline{\mathbf{R}}_{dd}$
as the signal subspace, and can then apply the ESPRIT
algorithm on this matrix.

\section{\label{sec:Related-Works}Related Sampling Problems}

In the introduction, we outlined previous approaches to time-delay
estimation. In this section, we explore in more detail the
relationship between our sampling problem and previous related
setups treated in the sampling literature: sampling signals from a union of subspaces
\cite{4483755,BD09},
compressed sensing of analog signals \cite{location,ME07,Mishali,EM082},
and FRI sampling \cite{1003065,4156380}.

\subsection{\label{sub:union}Sampling signals from a union of subspaces}
A signal model which received growing interest recently is that of signals
that lie in a union of subspaces \cite{4483755,BD09,EM082,ME07,Mishali,E082}.
Under this model each signal $x\left(t\right)$ can be described as \cite{4483755}
\begin{equation}
x\left(t\right) \in \bigcup_{\gamma \in \Gamma} \mathcal{S}_{\gamma},
\label{eq:union}
\end{equation}
where $\mathcal{S}_{\gamma}$ are subspaces of a given Hilbert space and $\Gamma$ is an index set.
The signal $x(t)$ lies in one of the subspaces $\mathcal{S}_{\gamma}$, however it is not known in advance in which one. Thus, effectively, to determine $x(t)$, we first need to find the active subspace, or the index $\gamma$.

Our signal model, given by
\eqref{eq:x(t)}, can be formulated as in \eqref{eq:union}.
As described in
Subsection \ref{subsec:Known} once the time delays are fixed, each signal $x\left(t\right)$
lies in a SI subspace spanned by $K$ generators. Therefore, the set of all signals
of the form \eqref{eq:x(t)} constitute an infinite union of SI subspaces, where $\gamma$ is the set of delays
$\tau$, which can take on any continuous value in the interval $[0,T]$, and $\mathcal{S}_{\gamma}$ is the corresponding SI subspace.

In \cite{4483755,BD09}
necessary and sufficient conditions are derived for a sampling operator to be
invertible over a union of subspaces. For the case of a union of SI
subspaces, \cite{4483755} suggests a sampling scheme, similar to that used in \cite{location} and in this paper,
comprised of parallel sampling channels.
Conditions on the sampling filters
are then given in order to ensure reconstruction of the signals from its samples.
In addition, the minimal number of sampling channels allowing perfect recovery of
the signal from its samples is shown to be $2K$. This leads to a minimal sampling
rate of $2K/T$ which is achieved by our scheme.
However, in \cite{4483755} no concrete reconstruction algorithms were given that can achieve this rate. Furthermore, although conditions for
invertibility were provided, these do not necessarily imply that there exists an efficient recovery algorithm, which can recover the signal from its samples at the minimal rate. Our aim in this work, is to provide concrete recovery techniques, that are simple to implement, for signals over an infinite union of SI subspaces.

In summary, in this work we focus on a special case of signals that lie in an infinite union of SI subspaces. For this case, in contrast to \cite{4483755}, we provide a concrete reconstruction method. This method achieves the minimal theoretical sampling rate derived in \cite{4483755}. In addition, while other works \cite{EM082,ME07,Mishali,location,E082} provided reconstruction algorithms only for signals defined over a finite union of subspaces, here we provide a first systematic sampling and reconstruction method for signals in an infinite union of subspaces.

\subsection{\label{sub:Compressed-sensing-of}Compressed sensing of analog signals}

The sampling problem in \cite{location} also deals with signals that
lie in a union of SI spaces and provides recovery algorithms. However in the setting of \cite{location} there are
finite number of possible subspaces, in contrast to our case, where
there are an infinite number of possible subspaces.

The signal model in \cite{location} can be described in terms of $N$ generating functions $a_{\ell}(t)$ as
\begin{equation}
x\left(t\right)=\sum_{\left|\ell\right|=K}\sum_{n\in\mathbb{Z}}d_{\ell}\left[n\right]a_{\ell}\left(t-nT\right),\label{eq:x(t)_cs}\end{equation}
where the notation $\left|\ell\right|=K$ means a sum over at most
$K$ elements. Thus, for each signal there are only $K$ active
generating functions out of $N$ total possible functions, but we
do not know in advance which generators are active. In principle,
such signals can be sampled and recovered using the paradigm
described in Section~\ref{sec:Sampling-Scheme} corresponding to
$N$ generating functions. Indeed, any signal of the form
(\ref{eq:x(t)_cs}) clearly also lies in the SI subspace spanned by
the $N$ generators $a_{\ell}(t)$, where some of the sequences
$d_{\ell}[n]$ are identically $0$. However, this would require a
sampling rate of $N/T$, obtained by $N$ sampling filters. Since
only $K$ of the generators are active, intuitively, we should be
able to reduce the rate and still be able to recover the signal.
The main contribution of \cite{location} is a sampling scheme
consisting of $2K$ filters that is sufficient in order to recover
$x(t)$ exactly.

We can formulate our problem as a finite union of SI spaces of the form
(\ref{eq:x(t)_cs}) if we assume that the $K$ unknown delays are
taken from a discrete grid containing $N$ possible time delays.
Under this assumption the generating functions in
\eqref{eq:x(t)_cs} can be expressed as
\begin{equation}
a_{\ell}\left(t\right)=g\left(t-t_{\ell}\right),\quad1\leq\ell\leq
N.\end{equation} Therefore, under a discrete setting, the method
of \cite{location} can provide a sampling and reconstruction
scheme for a signal of the form \eqref{eq:x(t)} with rate $2K/T$.

Similar to our approach here, the sampling scheme in
\cite{location} is based on $2K$ parallel channels, each comprised
of a filter and a uniform sampler at rate $1/T$. However, in order
to achieve this minimum sampling rate, the reconstruction in
\cite{location} involves brute-force solving an optimization
problem with combinatorial complexity. The complexity of the
reconstruction stage can be reduced by increasing the number of
channels, which entails a price in terms of  sampling rate. In
contrast, our reconstruction algorithm is based on the ESPRIT
algorithm, and can obtain the minimal sampling rate of $2K/T$ in
polynomial complexity. Furthermore, we do not require
discritization of the time delays but rather can accommodate any
continuous set of delays. In this sense we can view our sampling
paradigm as a special case of compressed sensing for an infinite
union of SI spaces. Since previous work in this area has focused
on sampling methods for finite unions, this appears to be a first
systematic example of a sampling theory where the subspace is
chosen over an infinite union.

Another difference with the approach of \cite{location} is the
design of the sampling filters. In our method, we have seen that
simple sampling filters can be used, such as low pass filter or bandpass filter-bank.
In contrast, the scheme of \cite{location} requires proper design of the sampling filters,
which is obtained in two stages. In the first stage, $N$ filters
$h_{\ell}\left(t\right),1\leq\ell\leq N$ are chosen that satisfy
some conditions with respect to the $N$ possible generating
functions. At the second stage, a smaller set of $p \geq 2K$
filters $s_i(t)$ is constructed from $h_{\ell}(t)$, via
\begin{equation}
s_{i}\left(t\right)=\sum_{\ell=1}^{N}\sum_{m=1}^{N}\sum_{n\in\mathbb{Z}}\mathbf{A}_{i\ell}^{*}c_{im}\left[n\right]h_{m}\left(t-nT\right),\label{eq:s_i(t)}\end{equation}
where $\mathbf{A}$ is a $p\times N$ matrix that satisfies the
requirements of compressed sensing in the appropriate dimension
\cite{donoho2006cs}, and $c_{im}\left[n\right]$ are a set of
sequences given explicitly in \cite{location}. In order to arrive
at filters that are easy to implement, a careful choice of the
parameters is needed, which may be difficult to obtain.

\subsection{Signals with finite rate of innovation}

Another interesting class of signals that has been treated
recently in the sampling literature are FRI signals
\cite{1003065,4156380}. Such signals have a finite number of
degrees of freedom per unit time, referred to as the rate of
innovation. Examples of FRI signals include streams of diracs,
nonuniform splines, and piecewise polynomials. A general form of
an FRI signal is given by \cite{1003065}\begin{equation}
x\left(t\right)=\sum_{n\in\mathbb{Z}}c_{n}\phi\left(t-t_{n}\right),\label{eq:x(t)_fri}\end{equation}
where $\phi\left(t\right)$ is a known function, $t_{n}$ are
unknown time shifts and $c_{n}$ are unknown weighing coefficients.
Recovery of such signals from their samples is equivalent to the
recovery of the delays $t_n$ and the weights $c_n$.

 Our signal
model \eqref{eq:x(t)} can be seen as a special case of
\eqref{eq:x(t)_fri}, where additional shift invariant structure is
imposed. This means that in each period $T$ the time delays are
constant relative to the beginning of the period, whereas in a
general FRI signal the time delay can vary from period to period.
Our method is designed in such a way that it utilizes this extra
structure to reduce the rate, while still guaranteeing perfect
recovery.

The FRI signals dealt with in \cite{1003065,4156380} are divided
into three main classes: periodic, finite length and infinite
length. If we address our signal model as an FRI signal it will
generally fall into the third category of infinite length FRI
signals. Some special classes of finite (and periodic) FRI signals
where treated in \cite{1003065}, such as streams of diracs. For
these special settings sampling theorems where derived with very
specific kernels, that achieve the minimal rate (the rate of
innovation). However, these methods are not adapted to the general
model (\ref{eq:x(t)_fri}).

Sampling and reconstruction of infinite length FRI signals was
treated in \cite{4156380}. The method in \cite{4156380} is based
on the use of specific sampling kernels which have finite time
support: kernels that can reproduce polynomials or exponentials.
In addition the function $\phi\left(t\right)$ is limited to
diracs, differentiated diracs, or short pulses with compact
support and no DC component. The reconstruction algorithm proposed
in \cite{4156380} is local, namely it recovers the signal's
parameters in each time interval separately. Naive use of this
approach in our context has two main disadvantages. First, in our
method the unknown delays are recovered from all the samples of
the signal $x\left(t\right)$. A local algorithm is less robust to
noise and does not take the shared information into account.  In
addition, in terms of  computational complexity, in our method all
the samples are collected to form a finite size correlation
matrix, and then the ESPRIT algorithm is applied once. Using the
local algorithm requires applying the annihilating filter method,
used for FRI recovery, on each time interval over again.

A final disadvantage of the FRI approach is the higher sampling
rate required. In order to discuss the sampling rate achieved by
the local algorithm proposed in \cite{4156380}, we limit our
discussion to the case where the function $\phi\left(t\right)$ is
a dirac, which is the main case dealt with in \cite{4156380}. The
theorems for unique recovery of the signal from its samples in
\cite{4156380} require that in each interval of size $2KLT_{s}$
there are at most $K$ diracs, where $L$ is the support of the
sampling kernel and $T_{s}$ is the sampling period. Since in each
interval of size $T$ we have $K$ diracs, it can be easily shown
that the minimal sampling rate is $2KL/T$, which is a factor of
$L$ larger than the rate achieved by our scheme. For example, when
using a B-spline kernel, which is the function with the shortest
time support that can reproduce polynomials of a certain order, an
order of at least $N=2K-1$ is needed, which has time support
$L=2K$. Thus, the sampling rate is $2K$ times larger than our
approach.

\section{\label{sec:Application}Application}

In this section we describe a possible application of the proposed
signal model and sampling scheme to the problem of channel
estimation in wireless communication \cite{meyr1997dcr}. In such
an application a transmitted communication signal passes through a
multipath time-varying channel. The aim of the receiver is to
estimate the channel's parameters from the samples of the received
signal.

We consider a baseband communication system operating in a
multipath fading environment with pulse amplitude modulation
(PAM). The data symbols are transmitted at a symbol rate of $1/T$,
modulated by a known pulse $g\left(t\right)$.
For this communication system the transmitted signal
$x_{t}\left(t\right)$ is given by\begin{equation}
x_{T}\left(t\right)=\sum_{n=1}^{N_{sym}}d\left[n\right]g\left(t-nT\right)\end{equation}
where $d\left[m\right]$ are the data symbols taken from a finite
alphabet, and $N_{sym}$ is the total number of transmitted
symbols.

The transmitted signal $x_{T}\left(t\right)$ passes through a
baseband time-varying multipath channel whose impulse response is
modeled as \cite{Proakis} \begin{equation}
h(\tau,t)=\sum_{k=1}^{K}\alpha_{k}\left(t\right)\delta\left(\tau-\tau_{k}\right)\label{eq:h(tau,t)}\end{equation}
where $\alpha_{k}\left(t\right)$ is the path time varying complex
gain for the $k$th multipath propagation path and $\tau_{k}$ is
the corresponding time delay. The total number of paths is denoted
by $K$. We assume that the channel is slowly varying relative to
the symbol rate, so that the path gains are considered to be
constant over one symbol period:\begin{equation}
\alpha_{k}\left(t\right)=\alpha_{k}\left[nT\right]\textrm{ for
}t\in\left[nT,\left(n+1\right)T\right].\end{equation} In addition,
we assume that the propagation delays are confined to one symbol,
i.e $\tau_{k}\in\left[0,T\right)$. Under these assumptions, the
received signal at the receiver is given by\begin{equation}
x_{R}\left(t\right)=\sum_{k=1}^{K}\sum_{n=1}^{N_{sym}}a_{k}\left[n\right]g\left(t-\tau_{k}-nT\right)+n\left(t\right)\label{eq:Xr(t)}\end{equation}
where \begin{equation}
a_{k}\left[n\right]=\alpha_{k}\left[nT\right]d\left[n\right]\end{equation}
and $n\left(t\right)$ denotes the channel noise.

The received signal $x_{R}\left(t\right)$ fits the signal model
described in \eqref{eq:x(t)}. Therefore, if the pulse shape
$g\left(t\right)$ satisfies the condition \eqref{eq:G(omega)} with
$p=2K$, our sampling scheme can recover the time delays of the
propagation paths. In addition, if the transmitted symbols are
known to the receiver, the time varying path gains can be
recovered from the sequences $a_{k}\left[n\right]$.

As a result our sampling scheme can estimate the channel's
parameters from samples of the output at a low rate, proportional
to the number of paths. As an example, we can look at the channel
estimation problem in code division multiple access (CDMA)
communication. This problem was handled using subspace techniques
in \cite{bensley1996sbc,strom1996dcs}. In these works the sampling
is done at the chip rate $1/T_{c}$ or above, where $T_{c}$ is the
chip duration given by $T_{c}=T/N$ and $N$ is the spreading factor
which is usually high ($1023$, for example, in GPS applications).
In contrast, our sampling scheme can provide recovery of the
channel's parameters at a sampling rate of $2K/T$. For a channel
with a small number of paths, this sampling rate can be
significantly lower than the chip rate.

Another example is UWB \cite{yang2004ultra} communications
which has gained popularity recently.
In this technology the bandwidth of the transmitted pulse can be up to several gigahertz.
Current technology commercial ADCs cannot operate at these sampling rates.
For example, the highest sampling rate ADC device, manufactured by National Semiconductor, supports sampling rates
 of up to $3$GHz at a relatively low resolution of $8$ bits and high power consumption.
In contrast, our proposed method, has a potential of reducing the sampling rate, into rates which can be achieved by lower rate ADCs with better resolution and lower power consumption.

\section{\label{sec:Numerical-Experiments}Numerical Experiments}
We now provide several experiments in which we examine various aspects of our proposed method.
The numerical experiments are divided into $4$ parts:
\begin{enumerate}
\item demonstration of a channel estimation application;
\item evaluation of performance in the presence of noise;
\item effects of approximation of the matrix $\mathbf{R}_{dd}$ using only a finite number of measurement vectors;
\item effects of imperfect digital correction filtering, using finite length filters.
\end{enumerate}

In all the simulations, except for the one in \ref{sub:DigFilter}, we use the sampling scheme described in
Section~\ref{sub:Complex-band-pass-filters}, which consists of a
bank of ideal band-pass filters. We assume that the working band
is $\mathcal{F}=\left[0,\frac{2\pi}{T}p\right]$, and that the
function $g\left(t\right)$ has constant frequency response on that
frequency range, i.e,
$G\left(\omega\right)=T,\omega\in\mathcal{F}$. In order to improve
the robustness to noise in the delay recovery stage, we use the
total least-squares (TLS) version of the ESPRIT algorithm
described in \cite{32276}. All the results are based on averaging
$1000$ experiments.

\subsection{\label{sub:ChannelEst}Channel estimation}
In the first simulation we demonstrate a channel estimation
application. We consider a time-varying channel of the form
\eqref{eq:h(tau,t)}, with $K=4$ paths. In order to simulate a time
varying channel, the channel's gain coefficients
$\alpha_{k}\left[nT\right]$ are modeled according to the Jakes'
model \cite{nla.cat-vn298883} as a zero-mean complex-valued
Gaussian stationary process with the classical U-shape power
spectral density. In such a model the varying rate of each gain
coefficient depends on the maximal Doppler shift $f_{d}$. In order
to simulate a slow varying channel, relatively to the symbol rate
$1/T$, we used for each path a maximal Doppler shift of
$f_{d}=0.05/T$. The energy of each time-varying path gain
coefficient was normalized to $\left(1/2\right)^{-k+1}$. The path
delays were drawn uniformly in the range $\left[0,T\right]$. For
the estimation $N_{sym}=100$ symbols were used where the data
symbols are assumed to be known. The samples at the output of each
of the sampling channels were corrupted by complex-valued Gaussian
white noise with an SNR of $15$dB.

The number of sampling channels is taken to be $p=5$, which is
only one more than the number of unknown delays. Although we have
seen that $2K$ sampling channels are required for perfect recovery
of every signal of the form \eqref{eq:x(t)}, for some signals
lowering the number of sampling channels is possible. Indeed,
according to Proposition \ref{pro:unique}, for signals with
$\textrm{dim}\left(\textrm{span}\left(\mathbf{b}\left[\Lambda\right]\right)\right)=K$,
the minimal number of sampling channels required is $K+1$. We will
demonstrate that for this example, $K+1$ channels are sufficient.

\begin{figure}
\begin{center}
\includegraphics[scale=0.5]{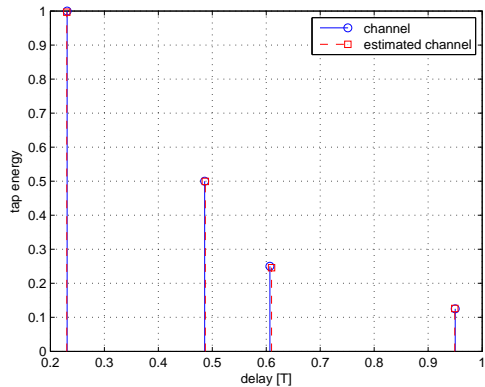}

\caption{\label{fig:PDP}Channel estimation with $p=5$ sampling
channels, and SNR=$15$dB.}
\end{center}
\end{figure}

\begin{figure}
\begin{center}
\includegraphics[scale=0.5]{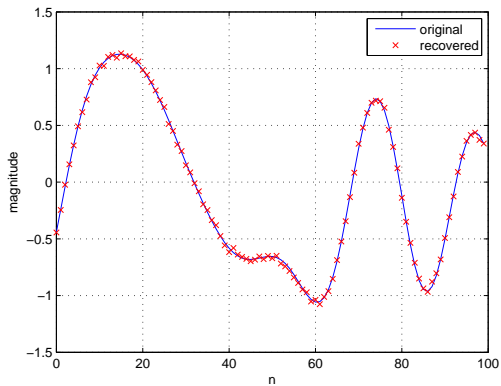}

\caption{\label{fig:FirstTap}Estimation of the time-varying gain coefficient
of the first path, $p=5$, SNR=$15$dB.}
\end{center}
\end{figure}

In Fig.~\ref{fig:PDP} the original and estimated channels are
shown. Since the gain coefficients of the channel are
time-varying, only their averaged energy over time is shown in the
figure. In Fig.~\ref{fig:FirstTap}, we plot the magnitude of the
original and estimated gains of the first path versus time. From
Figs.~\ref{fig:PDP} and \ref{fig:FirstTap} it is evident that our
method can provide a good estimate of the channel's parameters,
even when the samples are noisy.

\subsection{\label{sub:Noise}Performance in the presence of noise}
In the next simulations we examine the effect of SNR and the
number of sampling channels on the error in the delays estimation.
We choose $K=2$ close delays, $t_{1}=0.4352T$ and $t_{2}=0.521T$.
The sequences $a_k[n],k=1,2,\, n=1,2,\ldots100$ are
finite length sequences with unit power chosen according to Jakes'
model with $f_{d}=0.05/T$.

Under the setting of the simulation, which consists of a pulse with constant frequency response and ideal band-pass filters, from \eqref{eq:Decomposition} it can be verified that the sampling sequences satisfy the following relation in the time domain
\begin{equation}
\mathbf{c}\left[n\right]=\mathbf{N}\left(\tau\right)\mathbf{b}\left[n\right],\quad
n\in\mathbb{Z}.\label{eq:c[n]}\end{equation}
The Cramer-Rao bound (CRB) for unbiased estimators of $\theta_k = \frac{2\pi}{T}t_k$ from the data $\mathbf{c}\left[n\right]$, was derived in \cite{Cramer} for this data model.
The TLS-ESPRIT algorithm, used for the delays estimation in our method, is known to be asymptotically unbiased \cite{TLS_Per}.
Experimentally we verified that, under the simulation setup, the bias of the delays estimation is low enough for SNRs above $15$dB. Therefore, in this range of SNRs, the CRB derived in \cite{Cramer} can give a lower bound on the MSE
of the delays estimation (up to factor of $\frac{2\pi}{T}$), assuming our specific sampling scheme.

\begin{figure}
\begin{center}
\includegraphics[scale=0.5]{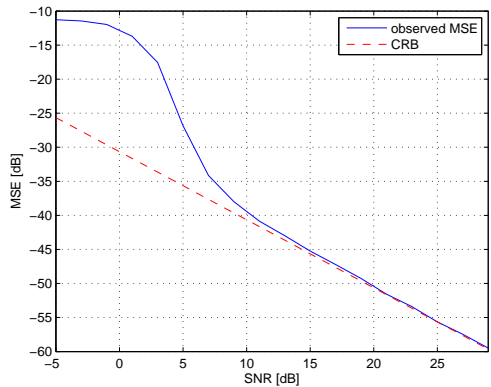}

\caption{\label{fig:VarSNR}MSE of the delays estimation versus
SNR, for $K=2$ and $p=4$.}
\end{center}
\end{figure}

In Fig.~\ref{fig:VarSNR}, the mean-squared error (MSE) of the
time delays estimation is shown versus the SNR, when using $p=4$
sampling channels. For comparison we also plot the CRB.
The figure demonstrates that our
method achieves the CRB for SNR$>15$dB, which is the range that delays estimation can be considered as unbiased.

\begin{figure}
\begin{center}
\includegraphics[scale=0.5]{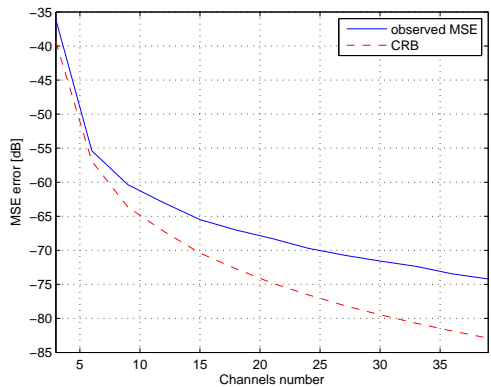}

\caption{\label{fig:VarP}MSE of the delays estimation versus the
number of sampling channels $p$, for $K=2$ and SNR=$10$dB.}
\end{center}
\end{figure}

In Fig.~\ref{fig:VarP}, the MSE of the estimation of the time delays is
shown versus the number of sampling channels, for a constant SNR
of $10$dB. The results demonstrate that the estimation error can be improved by increasing the number of
channels. Therefore, oversampling improves the robustness of our method to noise.

\subsection{\label{sub:RddEst}Effects of imperfect approximation of $\mathbf{R}_{dd}$}

\begin{figure}
\begin{center}
\includegraphics[scale=0.5]{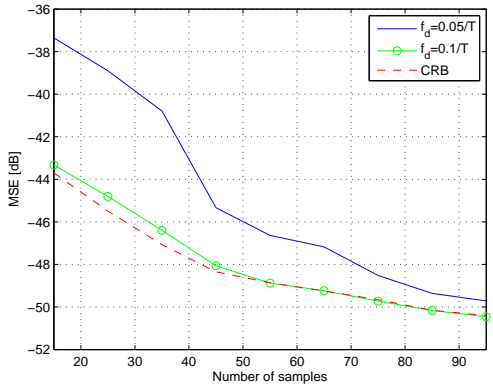}

\caption{\label{fig:VarN}MSE of the delays estimation versus the
number of samples used, for $K=2$, $p=4$ and SNR=$20$dB.}
\end{center}
\end{figure}

Next, we investigate the influence of estimating the matrix $\mathbf{R}_{dd}$
using only a finite number of measurement vectors $\mathbf{d}\left[n\right]$.
This number effects the total delay of our method, since reconstruction of the
sequences $a_k\left[n\right]$ is performed only after the unknown delays $\tau$ are recovered.
In Fig.~\ref{fig:VarN} the MSE of the delays estimation is shown versus the number of
measurement vectors used for estimation of $\mathbf{R}_{dd}$. A constant SNR of $20$dB and
$p=4$ sampling channels are used. Two cases are illustrated: in the first, the sequences
$a_k\left[n\right]$ are taken according to the Jakes' model with parameter $f_d=0.05/T$ and
in the second case $f_d=0.1/T$ is used, which corresponds to sequences with faster variation rate.
Fig.~\ref{fig:VarN} demonstrates that the MSE depends on the variation rate of the sequences. Intuitively the faster
the sequences vary, the more information each new measurement vector $\mathbf{d}\left[n\right]$ contains,
improving the estimation of $\mathbf{R}_{dd}$. In addition, it can be seen
that using only $50$ measurement vectors, yields a reasonable estimation of the delays in the case of $f_d=0.1/T$. The same estimation error is achieved using $80$ measurement vectors, when using slow varying sequences.
This result can be further improved by increasing the SNR or the number of sampling channels.

\subsection{\label{sub:DigFilter}Effects of imperfect digital filtering correction}
In the next simulation we examine the effects of approximating the digital correction
filter bank $\mathbf{W}^{-1}\left(e^{j\omega T}\right)$ by finite
length digital filters. The length of the filters affects the delay of our scheme.
To demonstrate this point, we arbitrarily choose a sampling scheme composed of $3$ non ideal
band-pass filters with a frequency response given by
\begin{equation}
S_{\ell}\left(\omega\right)=\begin{cases}
1.1-\left(1-0.4\ell\right)\textrm{cos}\left(\omega-\frac{2\pi}{T}\ell\right) & \omega\in \mathcal{F}_\ell\\
0 & \textrm{otherwise},\end{cases}
\end{equation}
where
\begin{equation}
\mathcal{F}_\ell=\left[\left(\ell-1\right)\frac{2\pi}{T},\ell\frac{2\pi}{T}\right].
\end{equation}
These filters satisfy the conditions of Proposition \ref{pro:d(omega)} and can model realistic sampling filters with non-flat frequency response. In this case a non trivial digital correction filter bank is required, whose coefficients are calculated using the inverse DTFT of $\mathbf{W}^{-1}\left(e^{j\omega T}\right)$.

\begin{figure}
\begin{center}
\includegraphics[scale=0.5]{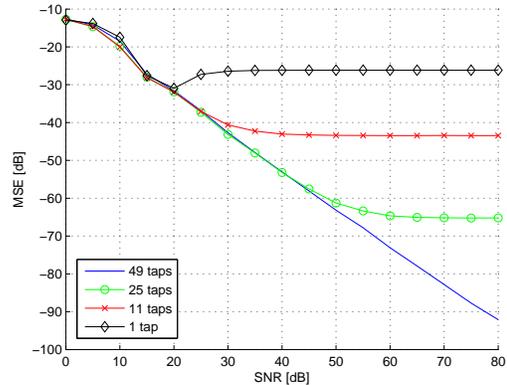}

\caption{\label{fig:VarLength}MSE of the delays estimation versus SNR
for different lengths of digital correction filter bank approximations.}
\end{center}
\end{figure}

In Fig.~\ref{fig:VarLength} the MSE of the delays estimation versus the SNR
is plotted for different lengths of filters.
At low SNRs the dominant error is caused by the noise, while for high SNRs the
error is mostly a result of the correction filter approximation. It can be seen that
a $49$ taps filters provide a good approximation to the correction filter bank,
resulting in a delay of $24$ samples. When working at SNRs below $40$dB, filters
with $11$ taps provide a reasonable approximation.

\section{Conclusion}

In this paper, we considered the problem of estimating the time
delays and time varying coefficients of a multipath channel, from
low-rate samples of the received signal. We showed that this
problem can be formulated within the broader context of sampling
theory, in which our goal is to recover an analog signal $x(t)$
that lies in a SI subspace, spanned by $K$ generators with unknown
delays. This class of problems can be viewed as an infinite union
of subspaces.

We showed that if the channel has $K$ multipath components, or
equivalently, if the SI subspace is generated by $K$ generators,
than under appropriate conditions on the sampling filters, a
sampling rate of $2K/T$ is necessary and sufficient to guarantee
perfect recovery of any signal $x(t)$. Here $T$ is the
transmission rate, or the period of the generators. We developed
 sufficient conditions on the generators and the
sampling filters in order to guarantee perfect recovery at the
minimal possible rate. To recover the unknown time delays, we
showed that our problem can be formulated within the context of
DOA estimation. Using this relationship, we proposed an
ESPRIT-type algorithm to determine the unknown delays from the
given low rate samples. Once the delays are properly identified,
the time varying coefficients can be found by digital filtering.

Besides the application to time delay estimation, the problem we
treated here can be seen as a first example of a systematic
sampling theory for analog signals defined over an infinite union
of subspaces. Recently, there has been growing interest in
sampling theorems for signals over a union of subspaces
\cite{4483755,BD09,EM082,ME07,Mishali,location,E082}. However,
previous work addressing stability issues and concrete recovery
algorithms have focused on finite unions. Here, we take a first
step in the direction of extending these ideas to a broader
setting that treats analog signals lying in an infinite union.

\section*{Acknowledgment}
The authors would like to thank the
anonymous reviewers for their valuable comments which helped improve the presentation.

\bibliographystyle{IEEEtran}
\bibliography{mutlipath}

\begin{thebibliography}{10}
\providecommand{\url}[1]{#1}
\csname url@samestyle\endcsname
\providecommand{\newblock}{\relax}
\providecommand{\bibinfo}[2]{#2}
\providecommand{\BIBentrySTDinterwordspacing}{\spaceskip=0pt\relax}
\providecommand{\BIBentryALTinterwordstretchfactor}{4}
\providecommand{\BIBentryALTinterwordspacing}{\spaceskip=\fontdimen2\font plus
\BIBentryALTinterwordstretchfactor\fontdimen3\font minus
  \fontdimen4\font\relax}
\providecommand{\BIBforeignlanguage}[2]{{%
\expandafter\ifx\csname l@#1\endcsname\relax
\typeout{** WARNING: IEEEtran.bst: No hyphenation pattern has been}%
\typeout{** loaded for the language `#1'. Using the pattern for}%
\typeout{** the default language instead.}%
\else
\language=\csname l@#1\endcsname
\fi
#2}}
\providecommand{\BIBdecl}{\relax}
\BIBdecl

\bibitem{quazi1981otd}
A.~Quazi, ``An overview on the time delay estimate in active and passive
  systems for target localization,'' \emph{IEEE Trans. on Acoustics, Speech and
  Signal Processing}, vol.~29, no.~3, pp. 527--533, 1981.

\bibitem{urick1983pus}
R.~J. Urick, \emph{{Principles of Underwater Sound}}.\hskip 1em plus 0.5em
  minus 0.4em\relax McGraw-Hill New York, 1983.

\bibitem{turin1980iss}
G.~L. Turin, ``Introduction to spread-spectrum antimultipath techniques and
  their application to urban digital radio,'' \emph{Proceedings of the IEEE},
  vol.~68, no.~3, pp. 328--353, March 1980.

\bibitem{bruckstein1985roe}
A.~Bruckstein, T.~J. Shan, and T.~Kailath, ``The resolution of overlapping
  echos,'' \emph{IEEE Trans. on Acoustics, Speech and Signal Processing},
  vol.~33, no.~6, pp. 1357--1367, 1985.

\bibitem{pallas1991ahr}
M.~A. Pallas and G.~Jourdain, ``Active high resolution time delay estimation
  for large {BT} signals,'' \emph{IEEE Trans. on Signal Processing}, vol.~39,
  no.~4, pp. 781--788, Apr 1991.

\bibitem{1143830}
R.~Schmidt, ``Multiple emitter location and signal parameter estimation,''
  \emph{IEEE Trans. on Antennas and Propagation}, vol.~34, no.~3, pp. 276--280,
  Mar 1986.

\bibitem{ziqiang1982nmh}
Z.~Q. Hou and Z.~D. Wu, ``A new method for high resolution estimation of time
  delay,'' \emph{IEEE International Conference on Acoustics, Speech, and Signal
  Processing, ICASSP '82}, vol.~7, pp. 420--423, May 1982.

\bibitem{saarnisaari1997tet}
H.~Saarnisaari, ``{TLS-ESPRIT} in a time delay estimation,'' \emph{IEEE 47th
  Vehicular Technology Conference, 1997}, vol.~3, pp. 1619--1623 vol.3, May
  1997.

\bibitem{4303294}
F.-X. Ge, D.~Shen, Y.~Peng, and V.~O.~K. Li, ``Super-resolution time delay
  estimation in multipath environments,'' \emph{IEEE Trans. on Circuits and
  Systems}, vol.~54, no.~9, pp. 1977--1986, Sept. 2007.

\bibitem{kumaresan1983eaa}
R.~Kumaresan and D.~W. Tufts, ``Estimating the angles of arrival of multiple
  plane waves,'' \emph{IEEE Trans. on Aerospace and Electronic Systems}, vol.
  AES-19, no.~1, pp. 134--139, Jan. 1983.

\bibitem{32276}
R.~Roy and T.~Kailath, ``{ESPRIT}-estimation of signal parameters via
  rotational invariance techniques,'' \emph{IEEE Trans. on Acoustics, Speech
  and Signal Processing}, vol.~37, no.~7, pp. 984--995, Jul 1989.

\bibitem{EM08}
Y.~C. Eldar and T.~Michaeli, ``Beyond bandlimited sampling,'' \emph{IEEE Signal
  Processing Magazine}, vol.~26, no.~3, pp. 48--68, May 2009.

\bibitem{Johnson1993}
D.~H. Johnson and D.~E. Dudgeon, \emph{\BIBforeignlanguage{English}{Array
  signal processing : concepts and techniques}}.\hskip 1em plus 0.5em minus
  0.4em\relax Englewood Cliffs, NJ: Prentice-Hall, 1993.

\bibitem{krim1996tda}
H.~Krim and M.~Viberg, ``Two decades of array signal processing research: the
  parametric approach,'' \emph{IEEE Signal Processing Magazine}, vol.~13,
  no.~4, pp. 67--94, Jul 1996.

\bibitem{4483755}
Y.~M. Lu and M.~N. Do, ``A theory for sampling signals from a union of
  subspaces,'' \emph{IEEE Trans. on Signal Processing}, vol.~56, no.~6, pp.
  2334--2345, June 2008.

\bibitem{BD09}
T.~Blumensath and M.~E. Davies, ``Sampling theorems for signals from the union
  of finite-dimensional linear subspaces,'' \emph{IEEE Trans. on Inform.
  Theory}, vol.~55, pp. 1872--1882, April 2009.

\bibitem{location}
Y.~C. Eldar, ``Compressed sensing of analog signals in shift-invariant
  spaces,'' \emph{IEEE Trans. on Signal Processing}, vol.~57, no.~8, pp.
  2986--2997, Aug. 2009.

\bibitem{ME07}
M.~Mishali and Y.~C. Eldar, ``Blind multiband signal reconstruction: Compressed
  sensing for analog signals,'' \emph{IEEE Trans. Signal Processing}, vol.~57,
  pp. 993--1009, Mar. 2009.

\bibitem{Mishali}
------, ``From theory to practice: Sub-{N}yquist sampling of sparse wideband
  analog signals,'' \emph{submitted to IEEE Selected Topics on Signal
  Processing}.

\bibitem{EM082}
Y.~C. Eldar and M.~Mishali, ``Robust recovery of signals from a structured
  union of subspaces,'' \emph{to appear in IEEE Trans. on Inform. Theory}.

\bibitem{4553693}
M.~Mishali and Y.~C. Eldar, ``Reduce and boost: Recovering arbitrary sets of
  jointly sparse vectors,'' \emph{IEEE Trans. on Signal Processing}, vol.~56,
  no.~10, pp. 4692--4702, Oct. 2008.

\bibitem{E082}
Y.~C. Eldar, ``Uncertainty relations for analog signals,'' \emph{to appear in
  IEEE Trans. on Inform. Theory}.

\bibitem{1003065}
M.~Vetterli, P.~Marziliano, and T.~Blu, ``Sampling signals with finite rate of
  innovation,'' \emph{IEEE Trans. on Signal Processing}, vol.~50, no.~6, pp.
  1417--1428, Jun 2002.

\bibitem{4156380}
P.~L. Dragotti, M.~Vetterli, and T.~Blu, ``Sampling moments and reconstructing
  signals of finite rate of innovation: Shannon meets strang-fix,'' \emph{IEEE
  Trans. on Signal Processing}, vol.~55, no.~5, pp. 1741--1757, May 2007.

\bibitem{AG01}
A.~Aldroubi and K.~Gr{\"o}chenig, ``Non-uniform sampling and reconstruction in
  shift-invariant spaces,'' \emph{{\em Siam Review}}, vol.~43, pp. 585--620,
  2001.

\bibitem{deboor1994sfg}
C.~de~Boor, R.~A. DeVore, and A.~Ron, ``The structure of finitely generated
  shift-invariant spaces in {$L_{2}\left(\mathbb{R}^{d}\right)$},''
  \emph{Journal of Functional Analysis}, vol. 119, no.~1, pp. 37--78, 1994.

\bibitem{geronimo1994ffa}
J.~S. Geronimo, D.~P. Hardin, and P.~R. Massopust, ``Fractal functions and
  wavelet expansions based on several scaling functions,'' \emph{Journal of
  Approximation Theory}, vol.~78, no.~3, pp. 373--401, 1994.

\bibitem{christensen2005gsi}
O.~Christensen and Y.~C. Eldar, ``Generalized shift-invariant systems and
  frames for subspaces,'' \emph{Journal of Fourier Analysis and Applications},
  vol.~11, no.~3, 2005.

\bibitem{stoica1997}
P.~Stoica and R.~Moses, \emph{Introduction to spectral analysis}.\hskip 1em
  plus 0.5em minus 0.4em\relax Englewood Cliffs, NJ: Prentice-Hall, 1997.

\bibitem{moulines1995}
E.~Moulines, P.~Duhamel, J.~Cardoso, S.~Mayrargue, and T.~Paris, ``Subspace
  methods for the blind identification of multichannel fir filters,''
  \emph{IEEE Trans. on Signal Processing}, vol.~43, no.~2, pp. 516--525, 1995.

\bibitem{32277}
M.~Wax and I.~Ziskind, ``On unique localization of multiple sources by passive
  sensor arrays,'' \emph{IEEE Trans. on Acoustics, Speech and Signal
  Processing}, vol.~37, no.~7, pp. 996--1000, Jul 1989.

\bibitem{uwb_char}
M.~Z. Win and R.~A. Scholtz, ``Characterization of ultra-wide bandwidth
  wireless indoor channels: {A} communication-theoretic view,'' \emph{IEEE
  Journal on Selected Areas in Communications}, vol.~20, no.~9, pp. 1613--1627,
  Dec 2002.

\bibitem{ADC}
B.~Le, T.~W. Rondeau, J.~H. Reed, and C.~W. Bostian, ``Analog-to-digital
  converters,'' \emph{IEEE Signal Processing Magazine}, vol.~22, no.~6, pp.
  69--77, Nov. 2005.

\bibitem{title}
J.~E. Evans, D.~F. Sun, and J.~R. Johnson, ``Application of advanced signal
  processing techniques to angle of arrival estimation in {ATC} navigation and
  surveillance systems,'' \emph{MIT, Lincoln Lab. Tech. Rep}, 1982.

\bibitem{1164649}
T.-J. Shan, M.~Wax, and T.~Kailath, ``On spatial smoothing for
  direction-of-arrival estimation of coherent signals,'' \emph{IEEE Trans. on
  Acoustics, Speech and Signal Processing}, vol.~33, no.~4, pp. 806--811, Aug
  1985.

\bibitem{donoho2006cs}
D.~L. Donoho, ``Compressed sensing,'' \emph{IEEE Trans. on Inform. Theory},
  vol.~52, no.~4, pp. 1289--1306, 2006.

\bibitem{meyr1997dcr}
H.~Meyr, M.~Moeneclaey, and S.~A. Fechtel, \emph{Digital communication
  receivers: synchronization, channel estimation, and signal processing}.\hskip
  1em plus 0.5em minus 0.4em\relax Wiley, New York, 1997.

\bibitem{Proakis}
J.~G. Proakis, \emph{\BIBforeignlanguage{English}{Digital communications}},
  3rd~ed.\hskip 1em plus 0.5em minus 0.4em\relax McGraw-Hill, New York, 1995.

\bibitem{bensley1996sbc}
S.~E. Bensley and B.~Aazhang, ``Subspace-based channel estimation for code
  division multiple access communication systems,'' \emph{IEEE Trans. on
  Communications}, vol.~44, no.~8, pp. 1009--1020, 1996.

\bibitem{strom1996dcs}
E.~G. Strom, S.~Parkvall, S.~L. Miller, and B.~E. Ottersten, ``{DS-CDMA}
  synchronization in time-varying fading channels,'' \emph{IEEE Journal on
  Selected Areas in Communications}, vol.~14, no.~8, pp. 1636--1642, Oct 1996.

\bibitem{yang2004ultra}
L.~Yang and G.~B. Giannakis, ``{Ultra-wideband communications: {An} idea whose
  time has come},'' \emph{IEEE Signal Processing Magazine}, vol.~21, no.~6, pp.
  26--54, 2004.

\bibitem{nla.cat-vn298883}
W.~C. Jakes, \emph{\BIBforeignlanguage{English}{Microwave mobile
  communications}}.\hskip 1em plus 0.5em minus 0.4em\relax Wiley, New York,
  1974.

\bibitem{Cramer}
P.~Stoica and N.~A., ``Music, maximum likelihood, and {Cramer-Rao} bound,''
  \emph{IEEE Trans. on Acoustics, Speech and Signal Processing}, vol.~37,
  no.~5, pp. 720--741, May 1989.

\bibitem{TLS_Per}
B.~Ottersten, M.~Viberg, and T.~Kailath, ``Performance analysis of the total
  least squares {ESPRIT} algorithm,'' \emph{IEEE Trans. on Signal Processing},
  vol.~39, pp. 1122--1135, May 1991.

\end{thebibliography}

\end{document}